\documentclass[aps,prb,twocolumn]{revtex4}

\usepackage{graphicx}

\begin{document}

\title{Scaling analysis of normal state properties of high-temperature
superconductors}

\author{H. G. Luo,$^1$ Y. H. Su,$^2$ and T. Xiang$^{3,1}$}

\address{$^{1}$Institute of Theoretical Physics, Chinese Academy of Sciences,
P.O. Box 2735, Beijing 100080, China}

\address{$^{2}$Department of Physics, Yantai University, Yantai 264005, China}

\address{$^{3}$Institute of Physics, Chinese Academy of Sciences, P.O. Box
603, Beijing 100080, China}

\begin{abstract}
We propose a model-independent scaling method to study the
physical properties of high-temperature superconductors in the
normal state. We have analyzed the experimental data of the c-axis
resistivity, the in-plane resistivity, the Hall coefficient, the
magnetic susceptibility, the spin-lattice relaxation rate, and the
thermoelectric power using this method. It is shown that all these
physical quantities exhibit good scaling behaviors, controlled
purely by the pseudogap energy scale in the normal state. The
doping dependence of the pseudogap obtained from this scaling
analysis agrees with the experimental results of angle-resolved
photoemission and other measurements. It sheds light on the
understanding of the basic electronic structure of high-$T_c$
oxides.
\end{abstract}

\maketitle

\section{Introduction}

The mechanism of high-$T_c$ superconductivity remains one of the
fundamental issues unsolved in condensed matter physics. In
particular, a unified theory toward the understanding of the rich
phase diagram of high-$T_c$ superconductors (HTSC) has not been
established. At half-filling, the parent compounds of HTSC are
antiferromagnetic insulators. Upon doping, the antiferromagnetic
long-range correlation is suppressed and the high-$T_c$
superconductivity develops above a critical doping level. At low
doping but in the metallic state, a pseudogap phase with missing
entropy or spectra is
discovered.\cite{PhysRevLett.63.1700,DingNature1996,LoeserShen1996,loram:1999}
Around the optimal doping, a strange metal phase with a linear
in-plane resistivity and probably a quantum critical point
emerges.\cite{Panagopoulos:2002} In the heavily overdoped regime,
the conventional Landau Fermi liquid behaviors are gradually
recovered. Throughout the whole doping range, the electron states
seem to be intrinsically inhomogeneous.\cite{Pan:2000}

A must step in the understanding of the phase diagram of HTSC is
to identify the energy scales and the control parameters or
interactions of low-energy excitations of HTSC. In the normal
state at low doping, the pseudogap is believed to be one of the
characteristic parameters of low-lying excitations. The
corresponding temperature below which the pseudogap effect is
observed is commonly used as a boundary to separate the pseudogap
and the strange metal phases, although there is no real phase
transition between these phases. However, due to the uncertainty
in the definition of the pseudogap, different experiments have
adopted different criteria to determine the pseudogap energy
scale. This leads to the claim of the existence of two pseudogaps,
namely, the upper and lower pseudogaps.\cite{timusk:1999} In the
strange metal phase, the quantum critical fluctuation is strong
and the temperature itself may serve as a dynamic control
parameter as suggested by the marginal Fermi liquid theory.
\cite{PhysRevLett.63.1996}

The scaling analysis of experimental data in HTSC is a simple but
powerful tool in elucidating the underlying physics without
invoking a specific model. In the normal state, if the pseudogap
is a predominant energy scale controlling low-energy excitations,
then the low-temperature behavior of any measurement physical
quantity should satisfy a doping-independent scaling law, although
the analytic expression of the scaling function is unknown. Based
on this idea, we analyzed recently the temperature dependence of
the c-axis resistivity $\rho_c$ of HTSC and found that it obeys a
universal scaling law given by\cite{su:134510}
\begin{equation} \label{sec1:rhoc}
\rho_c (T) = \frac{\alpha T}{\Delta}
\exp\left(\frac{\Delta}{T}\right),
\end{equation}
where $\alpha$ is a doping dependent coefficient, $T$ is
temperature, and $\Delta$ is the pseudogap. As shown in Ref.
\onlinecite{su:134510}, Eq. (\ref{sec1:rhoc}) results from the
interplay between the anisotropic c-axis hopping
integral\cite{PhysRevLett.77.4632,IntJModPhysB.12.1007} and the
$d_{x^2-y^2}$-like symmetry of the pseudogap. It agrees
excellently with the experimental data of multilayer HTSC and
resolved a long-standing puzzle regarding the physical origin of
the semiconductorlike temperature dependence of $\rho_c$ in the
pseudogap phase. Furthermore, it suggests that there is only one
energy scale controlling the low-energy excitations around the
antinodal points and the interlayer hopping within each unit cell
is coherent.

However, if the scaling function is unknown, the scaling analysis
becomes difficult. This has, in fact, hampered the application of
the scaling analysis. A commonly adopted approach in the scaling
analysis in the literature is to assume that by normalizing both
the measurement quantity $F(T)$ and the temperature by the
corresponding values at a sample-dependent characteristic
temperature $T^*$, then all the experimental data should fall onto
a single curve described by the scaling function
\begin{equation} \label{sec1:scaling}
\frac{F (T) }{F (T^{*})} = g \left(\frac{T}{T^{*}}\right).
\end{equation}
However, in real materials, this scaling analysis often fails
since $F(T)$ generally contains the terms which are not scaling
invariant. For example, the impurity contribution to the in-plane
resistivity or other physical quantities is not scaling variant.
Moreover, since $T^{*}$ is unknown prior to the analysis, this
formula is difficult to be implemented practically even if it is
correct. Empirically, $T^{*}$ is often determined from some
special features appearing in the measurement data. For example,
for the in-plane resistivity, $T^{*}$ is determined from the
temperature below which the resistivity begins to deviate from its
high-temperature linear-$T$ behavior. However, to determine
unambiguously the linear-$T$ region is not always possible since
in some low doping samples, the measured temperature may not be
high enough to reach the linear-$T$ regime. Furthermore, the
deviation from linear to nonlinear $T$ is a crossover, not a phase
transition, and a small measurement error may result in a large
error in $T^{*}$.

In this paper, we propose a scaling method and apply it to analyze
the normal state properties of HTSC. This method extends our
previous scaling analysis of the c-axis resistivity to other
physical quantities. It breaks the barrier in the use of the
simple formula (\ref{sec1:scaling}) and allows a model-independent
scaling analysis to be done reliably. We have reanalyzed the
experimental data of the c-axis resistivity of HTSC with this
method. By comparison with our previous results,\cite{su:134510}
we find that this method, indeed, works very well. It provides a
simple but powerful approach for analyzing experimental results.
Furthermore, this method is model independent. It can be applied
not only to the high-$T_c$ cuprates, but also to any other
materials where the single parameter scaling behavior is valid.

This paper is arranged as follows. Section \ref{sec:2} gives an
introduction to the scaling method based on a least square fit of
an unknown scaling function to the experimental data. In Sec.
\ref{sec:3}, we apply the method to analyze the scaling behaviors
of a number of physical quantities of HTSC in the normal state,
using the experimental data published in the literature. In Sec.
\ref{sec:4}, we analyze the universal behavior of the energy
scales obtained in Sec. \ref{sec:3} and discuss its physical
implications. Section \ref{sec:5} gives a brief summary.

\section{Method of scaling analysis}\label{sec:2}

In this section, we present a generic method for analyzing the
scaling behavior of a set of experimental data $\{F_i(T), i=1,...
,N\}$. Here, $T$ can be temperature, pressure, external field, or
any other controllable variable used in experiments. The subscript
$i$ is a sample index to which the physical quantity $F$ as a
function of $T$ is measured experimentally. $N$ is the total
number of samples. In the discussion below, in order to be
directly relevant to the scaling analysis presented in Sec.
\ref{sec:3}, we assume that $T$ is the temperature and a sample
represents a specified high-$T_c$ compound.

We start by assuming that in a relevant temperature range the
low-lying physics is governed only by one energy scale $\Delta$.
Thus the measured physical quantity $F(T)$ satisfies a simple
scaling law
\begin{equation}
F (T) =   \alpha  {\cal F} \left( \frac{T}{\Delta}\right) + \beta,
\label{sec2:eq1}
\end{equation}
where $\alpha$, $\beta$, and $\Delta$ are all doping dependent,
but temperature independent. For the physical quantities to be
discussed in Sec. \ref{sec:3}, $\beta$ is generally the
contribution of impurities or other extrinsic interactions.
$\Delta$ is a characteristic energy scale of the system. It
controls the dynamics of the system. ${\cal F}(T)$ is a universal
(doping-independent) scaling function. Its temperature dependence
is determined by the low-lying excitations and interactions.

The scaling function ${\cal F}(T)$ is generally unknown. This is the
difficulty commonly met in the data analysis. However, the scaling
method introduced here does not depend on the detailed formula of
${\cal F}(T)$, provided that the single-parameter scaling hypothesis
Eq. (\ref{sec2:eq1}) is valid. This is a merit of the method. It
provides a simple but powerful approach to probe the intrinsic
connection between different samples and to determine the doping
dependence of the characteristic energy scale $\Delta$ without
invoking a specific model.

The aim of the scaling analysis is to determine from the
measurement data the scaling parameters $(\alpha_i, \Delta_i ,
\beta_i)$ and the optimized scaling functions ${\cal F} (x)$ so
that all the data can be rescaled onto a universal curve. This
can, in principle, be achieved by minimizing the total deviation
of the scaling function between any two samples:
\begin{eqnarray}
\delta {\cal F} = \sum_{i<j}^{N} \sum_{k}^{N_{k}} \left[ {\cal
F}_i (T'_k) - {\cal F}_j (T'_k) \right] ^{2}, \label{sec2:eq2}
\end{eqnarray}
where
\begin{equation}
{\cal F}_i (T) = \frac{1}{\alpha_i }\left[F_i ( \Delta_i T ) -
\beta_i\right].  \label{sec2:eq3}
\end{equation}
$F_i(\Delta_i T)$ is the value of $F$ of the $i$th sample at
temperature $ \Delta_i T$. $N_k$ is the number of sampled
temperature points used in optimizing the scaling function. $N_k$
can be adjusted in the minimization. Initially, $N_k$ can take
roughly the value of average measured temperature points.

However, due to the scaling behavior of ${\cal F}(T)$, not all the
parameters $( \alpha_i, \Delta_i, \beta_i )$ can be uniquely
determined by the minimization of $\delta F$ if ${\cal F}$ is
unknown. Indeed, from Eqs. (\ref{sec2:eq2}) and (\ref{sec2:eq3}),
it can be shown that if $\{ \alpha_i, \Delta_i, \beta_i \}$ ($i =
1,... ,N$) is a set of parameters minimizing $\delta {\cal F}$,
then $( \alpha_i / \alpha_s , \Delta_i / \Delta_s , \beta_{i} -
\beta_s \alpha_i /\alpha_s )$ with arbitrary but nonsingular
$(\alpha_s, \Delta_s, \beta_s)$ will also minimize $\delta {\cal
F}$. $(\alpha_s, \Delta_s, \beta_s)$ are unknown and can be taken
as the scaling parameters of a reference sample. This means that
only the relative values of $\alpha_i$, $\Delta_i$, and
$\beta_{i}$ with respect to a reference sample
\begin{eqnarray}
A_{i}&=& \frac{\alpha_{i}}{\alpha_{s}}, \label{sec2:eq4} \\
B_{i}&=& \frac{\Delta_{i}}{\Delta_{s}}, \label{sec2:eq5} \\
C_{i} & = & \beta_i - A_i \beta_s , \label{sec2:eq6}
\end{eqnarray}
can be fixed. From the definitions
(\ref{sec2:eq4})-(\ref{sec2:eq6}), it is straightforward to show
that $A_s = B_s = 1$ and $C_s = 0$.

Using the relative scaling parameters $(A_i,B_i,C_i)$, one can
define a new scaling function
\begin{equation}
f \left(\frac{T}{B} \right) = \alpha_s {\cal F}\left(
\frac{T}{B\Delta_s} \right) + \beta_s .
\end{equation}
$F(T)$ can then be expressed as
\begin{equation}
F (T) =   A\, f \left( \frac{T}{B}\right) + C  . \label{sec2:eq1b}
\end{equation}
For the reference sample, the scaling function is the measurement
curve itself: $F (T) = f(T)$.

Equation (\ref{sec2:eq1b}) is nothing but to scale all
experimental data onto the measurement curve of the reference
sample. This suggests that the reference sample should be chosen
such that its temperature interval is broad enough to cover the
whole temperature range physically interesting and the data
quality is among one of the best.

The relative scaling parameters $\{A_i, B_i, C_i\}$ can now be
determined by minimizing the total deviation of the scaling
functions:
\begin{eqnarray}
\delta f = \sum_{i<j}^{N} \sum_{k}^{N_{k}} \left[ f_i (T_k) - f_j
(T_k) \right] ^{2}, \label{sec2:eq7}
\end{eqnarray}
where
\begin{equation}
f_i (T) = \frac{1}{A_i }\left[ F_i ( B_i T ) - C_i \right] ,
\label{sec2:eq8}
\end{equation}
In general, $ B_i T$ may not be exactly the temperature point
experimentally measured. The value of $F_i(B_i T)$ can be obtained
from the measurement data by interpolation, provided that $B_i T$
is within the measured temperature interval. For each pair of
$f_i$ and $f_j$, if $B_i T_k$ or $B_j T_k$ is outside the measured
temperature interval for the $i$th or $j$th sample, then the
corresponding term in Eq. (\ref{sec2:eq2}) should be excluded from
the summation. The minimization of $\delta f$ can be done, for
example, using the standard subroutine given in Ref.
\onlinecite{recipe}.

\begin{figure}[h]
\includegraphics[width=9cm]{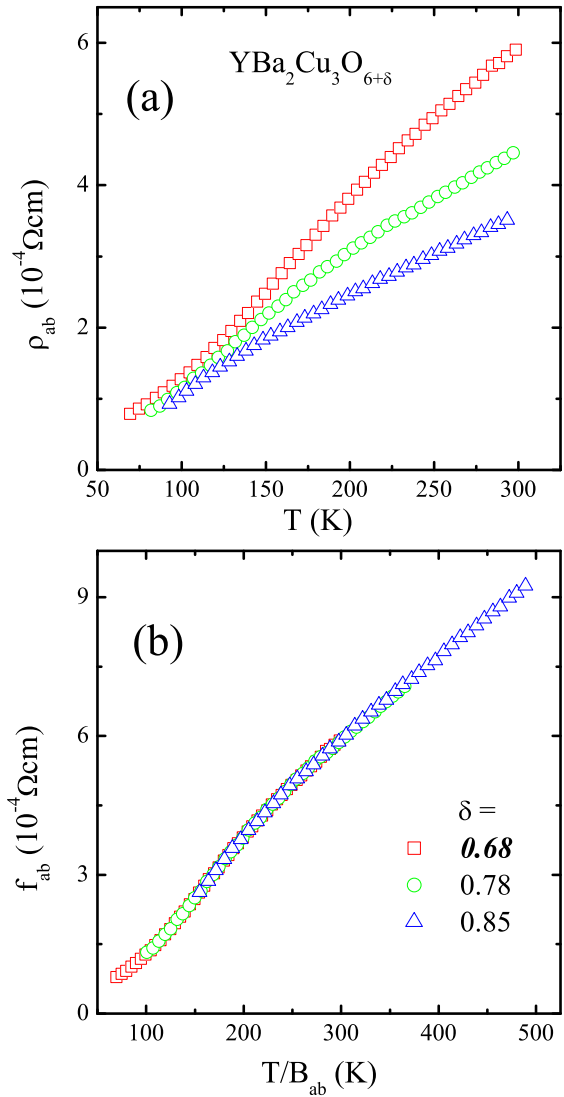}
\caption{(a) The experimental data of the in-plane resistivity
$\rho_{ab}(T)$ of YBa$_2$Cu$_3$O$_{6+\delta}$, extracted from Fig.
2(a) of Ref.~\onlinecite{PhysRevLett.70.3995}. (b) The scaling
plot of the experimental data. The reference sample is $\delta =
0.68$. } \label{app-fig}
\end{figure}

\begin{table}[h]
\caption{The scaling parameters for the in-plane resistivity
$\rho_{ab}$ of YBa$_2$Cu$_3$O$_{6+\delta}$ published by Ito
\textit{et al}.\cite{PhysRevLett.70.3995} The parameters of the
reference sample,  $\delta = 0.68$, are in bold face. }
\label{tab-app}
\begin{ruledtabular}
\begin{tabular}{llll}
Sample ($\delta$) &  $A$ & $B$ & $C$ ($10^{-4} \Omega$cm) \\
\hline
{\bf 0.68} & {\bf 1.0} & {\bf 1.0} & {\bf 0.0} \\
0.78 & 0.63 & 0.81 &  0.0074\\
0.85  & 0.39 & 0.60 &  -0.10
\end{tabular}
\end{ruledtabular}
\end{table}

Below we take the in-plane resistivity data of
YBa$_2$Cu$_3$O$_{6+\delta}$ published by Ito \textit{et
al}.\cite{PhysRevLett.70.3995} as an example to show how the
method works. For simplicity, here we only use the experimental
data for three of the samples, $\delta = (0.68, 0.78, 0.85)$.
Figure \ref{app-fig}(a) shows the measurement data of
$\rho_{ab}(T)$ for these three samples. In the scaling analysis,
we take $\delta = 0.68$ as the reference sample and $N_k$ to be
roughly equal to the measured temperature points of the reference
sample. The relative scaling parameters can then be determined by
numerically minimizing Eq. (\ref{sec2:eq7}). The resulting scaling
curves are shown in Fig. \ref{app-fig}(b) and the scaling
parameters are given in Table \ref{tab-app}.

In Eq. (\ref{sec2:eq1b}), if $C \ll F(T)$, then the scaling equation
is approximately given by
\begin{equation}
F(T) \approx A f\left( \frac{T}{B} \right).
\end{equation}
In this case, the scaling equation can be reexpressed as
\begin{equation} \label{sec2:sca3}
\frac{F(T)}{F(T^\prime)} =  \frac{f(T/B)}{f(T^* /B) } \equiv
g\left( \frac{T}{T^*} \right),
\end{equation}
where $T^\prime$ is a sample-dependent characteristic temperature.
$g(T)$ is a rescaled function of $f(T)$. Equation
(\ref{sec2:sca3}) is precisely the scaling equation defined by Eq.
(\ref{sec1:scaling}). It is a commonly used scaling equation in
the scaling analysis of experimental data. However, it should be
pointed out that this equation is valid only when the temperature
independent term $C$ can be safely ignored or reliably subtracted
from $F(T)$ in Eq. (\ref{sec2:eq1b}).

\section{Scaling analysis of experimental data} \label{sec:3}

In this section, we apply the scaling method to analyze the
experimental data of high-$T_c$ cuprates in the normal state. We
will first analyze the scaling behavior of the c-axis resistivity
$\rho_c$. Since an approximate but accurate expression for the
scaling function of $\rho_c$ is available, this allows us to
determine the absolute values of the scaling parameters. For other
measurement quantities, including the in-plane resistivity
$\rho_{ab}(T)$, the Hall coefficient $R_H(T)$, the magnetic
susceptibility $\chi(T)$, the spin-lattice relaxation rate
$1/T_1T$, and the thermoelectric power $S(T)$, only the relative
scaling parameters can be determined.

Our scaling analysis is based on the experimental data already
published in the literature. We collect as much as we can the
experimental data of HTSC from which a systematical analysis of
the scaling behaviors can be done. The preference is given to the
latest published data if there are considerable differences
between the data published by different groups. The chemical
formula of the compounds with their abbreviations analyzed in this
paper are given in Table \ref{tab:Abbrev}. The scaling analysis
here will be limited to the superconducting samples in the normal
state. The data for the nonsuperconducting samples will not be
analyzed. In all figures and tables presented in this paper, the
parameters for the reference samples will be in bold face to
distinguish them from other parameters.

\begin{table}[h]
\caption{ HTSC compounds and their abbreviations analyzed in this
paper. } \label{tab:Abbrev}
\begin{ruledtabular}
\begin{tabular}{ll}
YBa$_{2}$Cu$_{3}$O$_{6+\delta}$ & Y123 \\
Y$_{0.8}$Ca$_{0.2}$Ba$_{2}$Cu$_{3}$O$_{6+\delta}$ & Ca-Y123 \\
YBa$_{2}$Cu$_{4}$O$_{8}$ & Y124 \\
Bi$_{2}$Sr$_{2}$CaCu$_{2}$O$_{8+\delta}$ & Bi2212 \\
Bi$_{2}$Sr$_{2}$Ca$_{2}$Cu$_{3}$O$_{10+\delta}$ & Bi2223 \\
La$_{2-x}$Sr$_{2-x}$CuO$_{4}$ & La214 \\
Bi$_{2}$Sr$_{2-x}$La$_{x}$CuO$_{6+\delta}$ & La-Bi2201 \\
HgBa$_{2}$CuO$_{4}$ & Hg1201 \\
TlSr$_{2}$CaCu$_{2}$O$_{7-\delta}$ & Tl1212 \\
\end{tabular}
\end{ruledtabular}
\end{table}

In the comparison of the scaling parameters for different families
of HTSC, we will use the superconducting transition temperature
$T_c$ and its empirical formula proposed by Presland \textit{et
al.},\cite{PhysicaC176.95}
\begin{equation}
\frac{ T_{c}}{T_{c,max}}=1-82.6 (p-0.16)^2 , \label{eq:Tctop}
\end{equation}
to determine the effective carrier concentration $p$. Here,
$T_{c,max}$ is the maximal superconducting transition temperature.
For La$_{2-x}$Sr$_{x}$CuO$_{4}$, the carrier concentration is
equal to the doping concentration of Sr ions, $p = x$.

\subsection{c-axis resistivity $\rho_c(T)$}

In the pseudogap phase of high-$T_c$ cuprates, the c-axis
resistivity $\rho_c$ behaves very differently from its in-plane
counterpart $\rho_{ab}$. Along the CuO$_2$ plane, $\rho_{ab}$
shows a metal-like temperature dependence. It decreases with
decreasing temperature. However, along the c-axis, $\rho_{c}$
behaves as a semiconductor. It increases with decreasing
temperature.

This dramatic difference between $\rho_c$ and $\rho_{ab}$ is not
what one might expect within conventional Fermi liquid theory. To
resolve this issue, a number of theoretical models based on the
dynamic confinement of charge-spin separated particles
\cite{PhysRevLett.60.132,PhysRevB.45.5001,PhysRevB.52.10561} or
the incoherent interlayer hopping \cite{PhysRevB.63.064518,
PhysRevB.45.12636} were proposed. Most of the theories predicted
that $\rho_c$ should diverge in a certain power law of $T$ at low
temperature.  However, it seems that none of these theories can
account quantitatively or even qualitatively the experimental
data.

The semiconductorlike behavior of $\rho_c$ results, as we recently
pointed out, \cite{su:134510} from the interplay between the
$d_{x^2-y^2}$-like pseudogap and the anisotropic c-axis hopping
integral. \cite{PhysRevLett.77.4632,IntJModPhysB.12.1007} $\rho_c$
contributes mainly from the quasiparticles around the antinodal
points. The nodal contribution is completely suppressed by the
interlayer hopping matrix elements. Since the pseudogap is a
prevailing energy scale governing the c-axis dynamics in the
pseudogap phase, it is natural to assume that the c-axis
resistivity satisfies a scaling law governed purely by the
pseudogap $\Delta$.

\begin{figure}[ht]
\includegraphics[width=9cm]{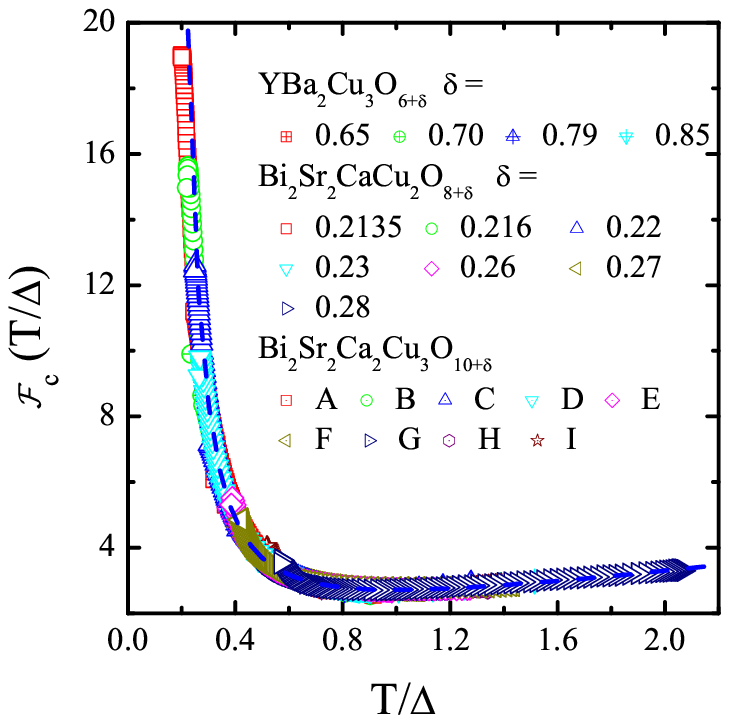}
\caption{The scaling function ${\cal F}_c$ of the c-axis
resistivity $\rho_{c}(T)$ for Y123, Bi2212, and Bi2223, compared
with the theoretical curve determined by Eq. (\ref{eq:rhoc2})
(dashed line). The experimental data are obtained from
Refs.~\onlinecite{PhysRevLett.79.2113, PhysRevLett.84.5848,
watanabe:note} for Bi2212, from
Ref.~\onlinecite{PhysRevB.66.024507} for Bi2223, and from Refs.
\onlinecite{PhysRevB.52.R751} and \onlinecite{PhysRevB.60.698} for
Y123. Bi2223 samples are labelled with the notations defined in
Ref.~\onlinecite{PhysRevB.66.024507}. } \label{fig:rhoc}
\end{figure}

\begin{figure}[ht]
\includegraphics[width=9cm]{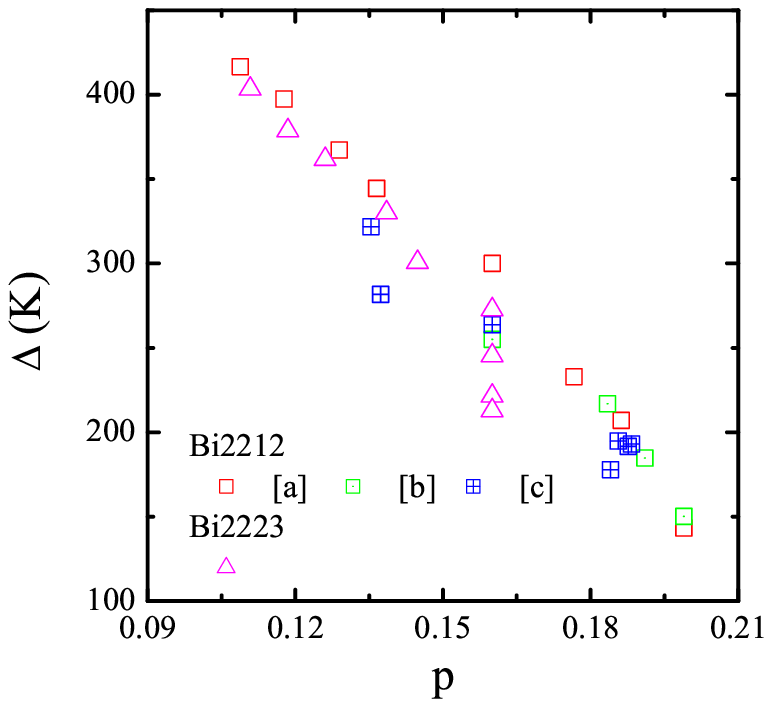}
\caption{The doping dependence of the pseudogap $\Delta$ obtained
from the scaling analysis of the experimental data shown in
Fig.~\ref{fig:rhoc} with Eqs. (\ref{eq:rhoc1}) and
(\ref{eq:rhoc2}). [a], [b], and [c] refer to Refs.
\onlinecite{PhysRevLett.79.2113}, \onlinecite{PhysRevLett.84.5848}
and \onlinecite{watanabe:note}, and
\onlinecite{PhysRevB.68.134505}, respectively. }
\label{fig:deltarc}
\end{figure}

In cuprate superconductors, if Cu atoms in the two neighboring
CuO$_2$ planes lie collinearly along the c-axis, then the
interlayer hopping integral between these two planes is given by
\begin{equation} \label{eq:tc}
t_c \sim (\cos k_x - \cos k_y)^2 ,
\end{equation}
where $(k_x, k_y)$ are the in-plane momenta of electrons. $t_c$
vanishes along the nodal direction. Based on this formula, we
showed in Ref. \onlinecite{su:134510} that for multilayer
cuprates, $\rho_c$ is approximately given by
\begin{equation} \label{eq:rhoc1}
\rho_c(T) = \alpha {\cal F}_c \left( \frac{T}{\Delta} \right),
\end{equation}
where
\begin{equation} \label{eq:rhoc2}
{\cal F}_c(x) = x \exp\left( \frac{1}{x} \right).
\end{equation}
Equation. (\ref{eq:rhoc1}) is a special case of Eq.
(\ref{sec2:eq1}). It holds when the residual resistivity
contributed by disorder scattering $\beta$ is vanishingly small
compared with the contribution of pseudogap to $\rho_c$.

The scaling function (\ref{eq:rhoc2}), as shown in Fig.
\ref{fig:rhoc}, agrees excellently with the measurement data for
Y123 published by Yan \textrm{et al}.\cite{PhysRevB.52.R751} and
by Babic \textrm{et al}. \cite{PhysRevB.60.698} for Bi2212 by
Watanabe \textrm{et
al}.\cite{PhysRevLett.79.2113,PhysRevLett.84.5848,watanabe:note}
and for Bi2223 by Fujii \textrm{et al}.\cite{PhysRevB.66.024507}

The values of the scaling parameters $\alpha$ and $\Delta$ are
given in Ref. \onlinecite{su:134510}. The pseudogap $\Delta$, as
shown in Fig. \ref{fig:deltarc}, drops almost linearly with
doping. This doping dependence of the pseudogap agrees with the
angle-resolved photoemission\cite{su:134510} (ARPES) as well as
other measurement data.\cite{Miyakawa:1998,tacon:2006} The values
of $\Delta$ (not shown here) for two overdoped samples of Bi2212
from Chen \textit{et al}.\cite{PhysRevB.58.14219} deviate
obviously from the other points. However, their $\rho_c$ data can
be well scaled onto the universal curve, as shown in Fig. 3(a) of
Ref. \onlinecite{su:134510}. A probable explanation is that the
true doping levels for these two samples may not be as high as
reported. The four samples, F, G, H and I, of Bi2223 near optimal
doping (here, we adopt the notations used in Ref.
\onlinecite{PhysRevB.66.024507}) have almost the same $T_c$.
However, their $\Delta$ obtained from $\rho_c$ are very different.
This might be due to the inhomogeneity of charge carriers in these
samples, since $T_c$ is determined mainly by the fraction of a
sample where the carrier concentration is close to the optimal
doping, while $\rho_c$ is the contribution of the whole sample.

The existence of the universal scaling law of $\rho_c$, especially
its activated behavior, implies that the c-axis hopping is
predominantly coherent, rather than incoherent as usually
believed. The reason is actually simple. If the interlayer hopping
is incoherent, then the excitations around the gap nodes may have
substantial contribution to $\rho_c$, which may break this scaling
law. For multiple layer cuprates, the intralayer coupling may be
different from the interlayer coupling. However, the different
coupling between CuO$_2$ planes does not change the fact that the
pseudogap is the only energy scale governing the quasiparticle
excitations around the antinodes in the normal state. Therefore,
Eq. (\ref{eq:rhoc1}) holds irrespective of the number of CuO$_2$
planes in each unit cell.

\begin{figure}[ht]
\includegraphics[width=9cm]{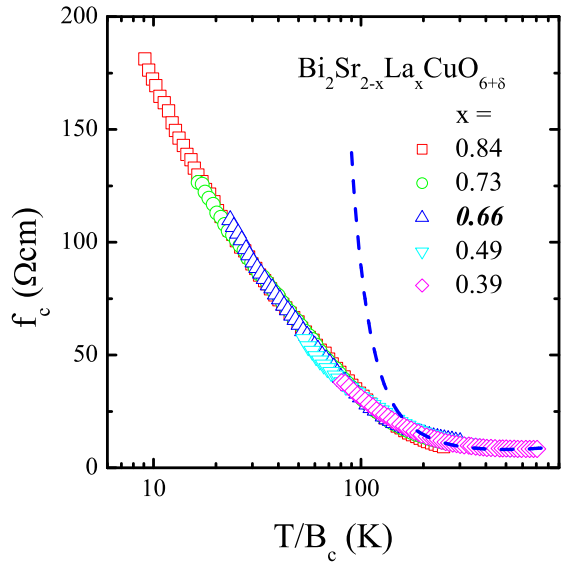}
\caption{Scaling behavior of the c-axis resistivity $\rho_{c}(T)$
for the single-layer cuprate superconductor La-Bi2201. The
experimental data were taken from
Ref.~\onlinecite{PhysRevB.67.104512}. The dashed line denotes the
scaling function defined by Eq. (\ref{eq:rhoc2}). }
\label{fig:BSLCO-rhoc}
\end{figure}

\begin{table}[h]
\caption{The scaling parameters, $A_c$ and $B_c$, of the c-axis
resistivity $\rho_c$ defined by Eq. (\ref{eq:rhoc3}) for
La-Bi2201. The experimental data and the carrier concentration p
are taken from Ref. \onlinecite{PhysRevB.67.104512}. }
\label{tab:rhoc2}
\begin{ruledtabular}
\begin{tabular}{lllll}
 Sample (x) & $T_c$ (K) & Doping (p) & $A_c$ & $B_c$ \\
 \hline
0.84 & 1.4 & 0.10 & 2.50 & 1.13 \\
0.73& 14 & 0.11 & 1.37 & 1.14 \\
{\bf 0.66} & {\bf 23} & {\bf 0.12} & {\bf 1.0} & {\bf 1.0} \\
0.49 & 31 & 0.14 & 0.79 & 0.51 \\
0.39& 38 & 0.16 & 0.67 & 0.41
\end{tabular}
\end{ruledtabular}
\end{table}

The above scaling analysis is done based on Eq. (\ref{eq:rhoc2})
by assuming that the interlayer hopping integral is given by Eq.
(\ref{eq:tc}). However, in single-layer cuprate compounds, Cu
atoms of two adjacent CuO$_2$ planes do not lie collinearly along
the c-axis. In this case, the c-axis hopping integral becomes
\cite{IntJModPhysB.12.1007}
\begin{equation}
t_c \sim (\cos k_x - \cos k_y)^2 \cos \frac{k_x}{2} \cos
\frac{k_y}{2}.
\end{equation}
It vanishes along both the nodal and antinodal directions. The
c-axis hopping, thus, contributes mainly from the low-lying
excitations between the nodal and antinodal points. As the
pseudogap is the dominant energy scale in the pseudogap phase, the
scaling law of the c-axis resistivity [Eq. (\ref{eq:rhoc1})]
should still hold. However, the scaling function of the
single-layer cuprates is expected to be different. In this case,
an accurate expression for the scaling function is not available.
Thus, we are unable to determine the absolute values of the
scaling parameters $\Delta$ and $\alpha$ from the scaling
analysis. However, the relative scaling parameters can be
determined using the scaling method introduced in the previous
section.

We have analyzed the scaling behavior of $\rho_c$ for La-Bi2201
(Ref. \onlinecite{PhysRevB.67.104512}) using the formula
\begin{equation} \label{eq:rhoc3}
\rho_c = A_c f_c \left(\frac{T}{B_c}\right) .
\end{equation}
Again, the contribution from impurity scattering to $\rho_c$ is
ignored since it is much smaller than the contribution from the
pseudogap effect.

Figure. \ref{fig:BSLCO-rhoc} shows the scaling function ${f}_c$
for La-Bi2201. The scaling parameters are given in Table
\ref{tab:rhoc2}. The analysis shows that $\rho_c$, indeed,
exhibits a good scaling behavior in this one-layer material.
However, the scaling function is different from that for the
multilayer materials at low temperatures. It is also different
from the logarithmic divergence behavior as observed by Ando
\textit{et al}. in La214 cuprates. \cite{PhysRevLett.75.4662}

The scaling behavior of $\rho_c$ for both the single- and
multiple-layer cuprates indicates that the interlayer dynamics is,
indeed, governed by the interplay between the anisotropic
interlayer hopping integral and the pseudogap effect in all
high-$T_c$ cuprates.

\begin{widetext}
\begin{center}
\begin{figure}[ht]
\includegraphics[width=16cm]{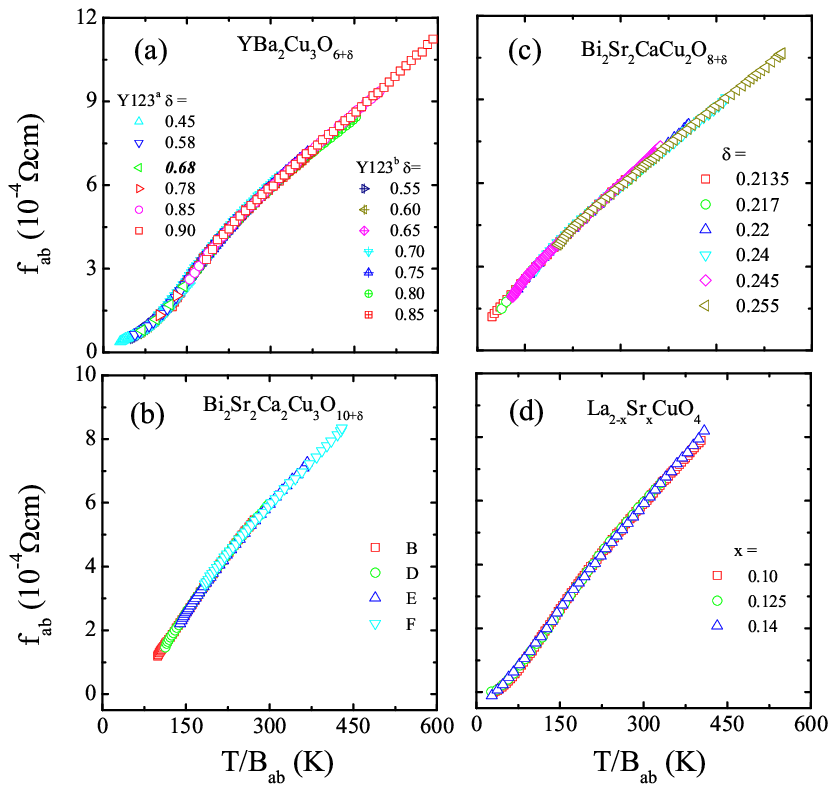}
\caption{The scaling function of the in-plane resistivity for
Y123$^{a}$ (Ref.~\onlinecite{PhysRevLett.70.3995}), Y123$^{b}$
(Ref.~\onlinecite{PhysRevLett.93.267001}), Bi2212
(Ref.~\onlinecite{PhysRevLett.79.2113}), Bi2223
(Ref.~\onlinecite{PhysRevB.66.024507}), and La214
(Ref.~\onlinecite{PhysRevB.49.16000}). The Bi2223 samples are
labeled using the notations given in
Ref.~\onlinecite{PhysRevB.66.024507}.} \label{fig:rhoab1}
\end{figure}
\end{center}
\end{widetext}

\subsection{In-plane resistivity $\rho_{ab}(T)$}
\label{in-plane}

In contrast to the semiconductorlike behavior of $\rho_c$, the
in-plane resistivity $\rho_{ab}$ of high-$T_c$ cuprate
superconductors is metal-like in the normal state. In the
underdoped and optimally doped materials, $\rho_{ab}$ exhibits a
universal linear behavior at high temperatures.
\cite{PhysicaC235.130} The phonon scattering can lead to a linear
resistivity. However, the Debye temperature determined by applying
the Bloch-Gruneisen formula is too low too account for the
experimental data  of Bi2201.\cite{PhysRevB.41.846} The linear
behavior of $\rho_{ab}$ could be a manifestation of strong
correlations. It is a characteristic behavior of marginal Fermi
liquid, \cite{PhysRevLett.63.1996} where the inelastic scattering
rate scales linearly with temperature. It may also result from
gauge\cite{PhysRevLett.64.2450,PhysRevLett.65.653,JPSJ59.2905} or
quantum critical fluctuations.

Below a characteristic temperature $T^*$, $\rho_{ab}$ begins to
deviate from the linear behavior. This deviation is correlated
with the pseudogap effect, and $T^*$ is believed to be the onset
temperature below which the pseudogap opens. However, the opening
of the pseudogap does not lead to a phase transition. There are no
thermal anomalies observed in the specific heat or other
thermodynamic quantities around $T^*$.

To analyze the scaling behavior of the in-plane resistivity, we
assume $\rho_{ab}$ to satisfy the following scaling law:
\begin{equation} \label{eq:rhoab}
\rho_{ab}(T) = A_{ab} f_{ab} \left( \frac{T}{B_{ab}}\right) + C_{ab}
.
\end{equation}
Here, the residual resistivity $C_{ab}$ should be retained since
the impurity contribution to $\rho_{ab}$ is no longer negligible
compared with the inelastic contribution of electrons to
$\rho_{ab}$.

\begin{table}[ht]
\caption{The scaling parameters, $A_{ab}$, $B_{ab}$, and $C_{ab}$
for the in-plane resistivity of Y123$^{a}$
(Ref.~\onlinecite{PhysRevLett.70.3995}), Y123$^{b}$
(Ref.~\onlinecite{PhysRevLett.93.267001}), Bi2212
(Ref.~\onlinecite{PhysRevLett.79.2113}), Bi2223
(Ref.~\onlinecite{PhysRevB.66.024507}), and La214
(Ref.~\onlinecite{PhysRevB.49.16000}) as shown in
Fig.~\ref{fig:rhoab1}. The notations used in Ref.
\onlinecite{PhysRevB.66.024507} are adopted to label the samples
of Bi2223. Y123$^{a}$ with $\delta=0.68$ is taken as the reference
sample. The unit of $C_{ab}$ is $10^{-4}\Omega$\,cm. The doping p
of Y123$^{a}$, Bi2212 and Bi2223 are obtained from the empirical
formula Eq.~(\ref{eq:Tctop}), with the maximum $T_c$ being $93.54$
K for $\delta=0.90$ of Y123$^{a}$, $89.0$ K for Bi2212 at
$\delta=0.22$, and $108.0$ K for F sample of Bi2223.  The values
of $T_{c}$ for La214, Y123$^{a}$, Bi2212, and Bi2223 are obtained
from Ref.~\onlinecite{loram:1999},
\onlinecite{PhysRevLett.70.3995},
\onlinecite{PhysRevLett.79.2113}, and
\onlinecite{PhysRevB.66.024507}, respectively. } \label{tab:rhoab}
\begin{ruledtabular}
\begin{tabular}{lllllll}
 & Sample & $T_{c}$ (K) & Doping (p)   & $A_{ab}$ & $B_{ab}$ & $C_{ab}$ \\
\hline
Y123$^{a}$ & 0.45 & 54.96 & 0.088  & 3.83& 1.91 & 0.09 \\
& 0.58 & 64.67 & 0.098  & 1.48 & 1.22  & 0.13\\
& {\bf 0.68} & {\bf 67.04} & {\bf 0.100} & {\bf 1.0} & {\bf 1.0} & {\bf 0.0} \\
& 0.78 & 80.26 & 0.118 & 0.63 & 0.81  & 0.0074\\
& 0.85 & 92.08 & 0.146  & 0.39 & 0.60 & -0.10 \\
& 0.90 & 93.54 & 0.16  & 0.22 & 0.50 & -0.19 \\
Y123$^{b}$ & 0.55 & &  & 2.02& 1.33 & 0.19 \\
& 0.60 & &  & 1.57 & 1.20 & -0.31 \\
& 0.65 & &  & 1.35 & 1.15 & -0.22 \\
& 0.70 & &  & 0.89 & 0.92 & -0.095 \\
& 0.75 & &  & 0.70 & 0.81 & -0.070\\
& 0.80 & &  & 0.55 & 0.65 & -0.26 \\
& 0.85 & &  & 0.52 & 0.78 & 0.049 \\
Bi2212 &0.2135 & 71.0 & 0.11 & 1.18& 0.96  & 1.70\\
&0.217& 77.0 & 0.119 & 1.07& 0.89 & 1.10 \\
&0.22 & 83.0 & 0.131 & 0.65& 0.73 & 1.12 \\
&0.24 & 89.0 & 0.16 & 0.51& 0.63 & 0.60 \\
&0.245 & 87.86 & 0.173  & 0.48& 0.82 & 0.76 \\
&0.255 & 87.4 & 0.175 & 0.32& 0.54 & 0.24 \\
Bi2223 &B & 93.0 & 0.118 & 0.12& 1.09 & 0.31 \\
& D & 104.0 & 0.139 & 0.10& 0.99 & -0.20 \\
& E & 106.0 & 0.145 & 0.087& 0.81 & -0.69 \\
&F & 108.0 & 0.16 & 0.07& 0.69  & -0.90\\
La214 & 0.10 &  30.75 & 0.10   & 0.84 & 2.06 &9.47\\
& 0.125 & 32.0 & 0.125  &0.60 &1.97 &6.93\\
&0.14 & 36.62 & 0.14  & 0.51 &2.05 & 5.77\\
\end{tabular}
\end{ruledtabular}
\end{table}

We have applied Eq.~(\ref{eq:rhoab}) to the experimental data
published by Ito \textrm{et al.}\cite{PhysRevLett.70.3995} for
Y123, by Watanabe \textrm{et al.}\cite{PhysRevLett.79.2113} for
Bi2212, by Fujii \textrm{et al.}\cite{PhysRevB.66.024507} for
Bi2223, and by Nakano {\it et al.} \cite{PhysRevB.49.16000} for
La214. Figure \ref{fig:rhoab1} shows the scaling function for
these compounds. The corresponding scaling parameters are shown in
Table \ref{tab:rhoab}. The measurement data (not shown in the
figure) begin to deviate from the universal scaling curves near
$T_c$ due to superconducting fluctuations.

The result of Fig. \ref{fig:rhoab1} shows that $\rho_{ab}$,
indeed, satisfies the simple scaling law described by Eq.
(\ref{eq:rhoab}). Moreover, all the curves shown in Fig.
\ref{fig:rhoab1} are obtained by taking Y123 $\delta = 0.68$ as a
reference. This means that $\rho_{ab}$ can be scaled onto a single
curve for all these materials. Thus the scaling function of
$\rho_{ab}$ is universal. It does not depend on the chemical
structure nor on the doping level. This suggests that the in-plane
resistivity is governed by the same scattering mechanism in all
cuprate superconductors

The striking scaling behavior of $\rho_{ab}$ indicates that the
characteristic temperature $T^*$ above which $\rho_{ab}$ varies
linearly with temperature is proportional to the scaling parameter
$B_{ab}$. In Ref. \onlinecite{su:134510}, we showed that $T^*$ is
proportional to the pseudogap $\Delta$ determined from the c-axis
resistivity within measurement errors, independent of doping.
Thus, $B_{ab}$ is also proportional to $\Delta$. This means that,
same as for the c-axis resistivity, the pseudogap $\Delta$ is also
a control energy scale for the in-plane resistivity, although
$\rho_{ab}$ is mainly the contribution of nodal quasiparticle
excitations.

Wuyts \textrm{et al}.\cite{PhysRevB.51.6115, PhysRevB.53.9418} did
a similar scaling analysis for the in-plane resistivity of Y123.
However, the characteristic temperatures (or energy scales) they
determined are not very accurate. Their scaling curves of
$\rho_{ab}$ (Fig.~1 of Ref.~\onlinecite{PhysRevB.51.6115} and
Fig.~7 of Ref.~\onlinecite{PhysRevB.53.9418}) do not look as good
as those shown in Fig. \ref{fig:rhoab1}.

In Fig.~\ref{fig:rhoab1}(d), we only show the experimental data
for slightly underdoped La214 samples ($x = 0.1, 0.125$, and
$0.14$). For heavily underdoped samples ($x = 0.04, 0.06$, and
$0.08$), we find that the data deviate significantly from the
universal scaling curve below $T^*$. This deviation was observed
only in the La214 samples. It may be due to the suppression of the
scattering rate by the formation of stripe or other competing
orders in these compounds.\cite{Nature375.561}

In the overdoped regime, $\rho_{ab}$ is not linear-$T$ dependent
in nearly the whole temperature range.\cite{PhysRevLett.69.2975}
The experimental data do not fall onto the scale curves as shown
in Fig.~\ref{fig:rhoab1}. This change of the temperature behavior
of $\rho_{ab}$ in the overdoped regime can be understood from the
change of the Fermi surface topology revealed by
ARPES.\cite{PhysRevB.65.094504} In the overdoped region, the Fermi
surface becomes electronlike. This may affect strongly the dynamic
behavior of electrons in the CuO$_2$ planes.

\begin{figure}[ht]
\includegraphics[width=7cm]{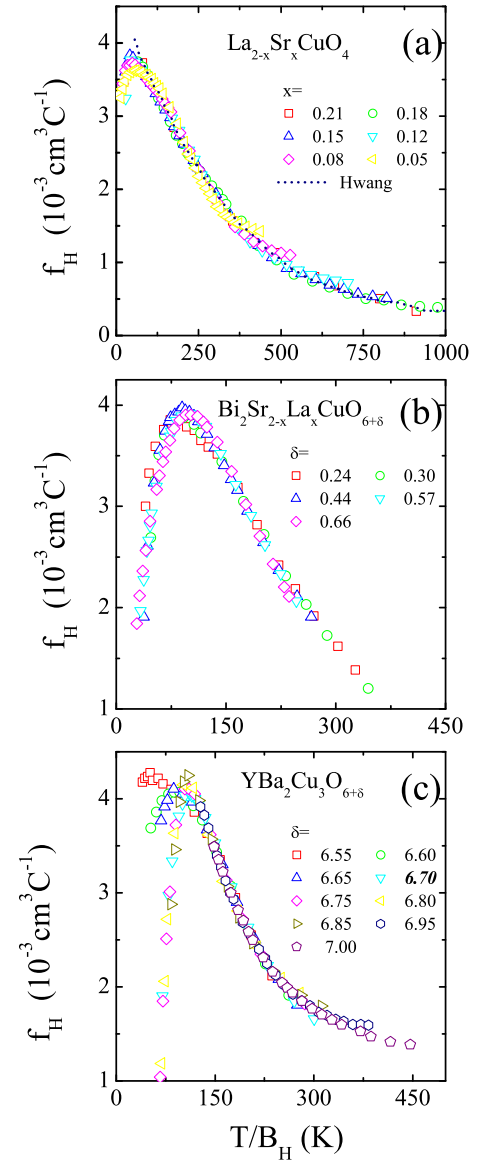}
\caption{The scaling functions of the Hall coefficient $R_H(T)$
for  (a) La214 (Ref.~\onlinecite{ono:024515}), (b) La-Bi2201
(Ref.~\onlinecite{ando:R6991}) and (c) Y123 (Ref.
\onlinecite{segawa:104521}). The dashed line in (a) is obtained
from the scaling curves determined by Hwang \textit{et al}.
(Ref.~\onlinecite{PhysRevLett.72.2636}). } \label{fig:Hall1}
\end{figure}

\subsection{Hall coefficient $R_H(T)$} \label{Hall}

The Hall coefficient $R_H(T)$ is an important quantity in
characterizing the nature of charge carriers. In a conventional
metal with spheric Fermi surface and isotropic scattering rate,
the Hall coefficient is inversely proportional to the carrier
concentration, independent of temperature. The sign of $R_H$
reflects the type of conducting charge carriers. $R_H$ is negative
or positive if the charge carriers are electrons or holes.
However, in doped transition metal oxides, such as high-$T_c$
cuprates, the Hall coefficient is strongly temperature
dependent.\cite{PhysRevLett.67.2088,
PhysRevB.42.8704,PhysRevB.46.8694} It is determined not just by
the carrier concentration, but also by the scattering rates and
the curvature of the Fermi surface. Other effects, such as
magnetic skew scattering, can also affect the temperature
dependence of $R_H$.\cite{PhysRevB.38.9198}

In HTSC, $R_{H}$ shows a strong temperature and doping dependence.
At high temperature,  $R_{H}$ increases rapidly with decreasing
temperature. After reaching a maximum, $R_H$ drops down to low
temperature in most of the samples. This is the typical
temperature dependence of $R_H$ in HTSC. It was observed in
Y123,\cite{PhysRevLett.67.2088, segawa:104521}
La214,\cite{PhysRevLett.72.2636, ando:197001, ono:024515} Bi
systems,\cite{ando:R6991, PhysRevB.62.R11989, fruchter:092502}
Hg1212,\cite{PhysRevB.50.3246} and Tl
systems.\cite{PhysRevB.47.1119, PhysRevB.50.6402,
PhysRevB.50.16033}

The complex temperature dependence of the Hall coefficient remains
one of the hardest problems to be resolved. Within the theory of
charge-spin separation, Anderson\cite{PhysRevLett.67.2092}
proposed to use the Hall angle $\Theta_{H}$ instead of the Hall
coefficient $R_H$ to understand the Hall anomaly. He argued that
due to the charge-spin separation, the Hall angle, which is
defined by the ratio between the transverse and longitudinal
conductivities, $\Theta_H = \tan^{-1} \sigma_{xy} / \sigma_{xx}$,
should be determined purely by the transverse relaxation rate
(i.e., the relaxation rate perpendicular to the Fermi surface).
This eliminates the ambiguity in the explanation of the Hall
coefficient, since it is determined by both the longitudinal and
transverse relaxation rates. Anderson further argued that as the
transverse relaxation rate is determined by the spin excitations,
which is relatively normal, the Hall angle should follow the
temperature dependence of normal Fermi liquid, i.e.,
\begin{equation}
\cot \Theta_{H} = \alpha T^{2} + \gamma,
\end{equation}
where $\alpha$ is a temperature independent coefficient and
$\gamma$ is the impurity contribution.\cite{PhysRevB.46.8687} This
quadratic temperature dependence of the Hall angle, indeed, agrees
with the experimental observation at optimal doping. However, in
both underdoped and overdoped regimes, \cite{segawa:104521,
ando:197001, PhysRevLett.72.2636, ando:R6991, PhysRevB.62.R11989}
the temperature exponent deviates generally from 2, and the above
expression breaks down.

\begin{table}[ht]
\caption{The scaling parameters for the Hall coefficient $R_H$
shown in Fig. \ref{fig:Hall1}. The unit of $C_H$ is $10^{-3}$
cm$^{3}$C$^{-1}$. T$_{c}$ for La214 are obtained by interpolating
the data given in Ref.~\onlinecite{loram:1999}. The hole
concentration p for La-Bi2201 and Y123 are obtained from
Eq.~(\ref{eq:Tctop}) with the maximum T$_{c,max} = 33.03$ K for
$x=0.44$ of La-Bi2201, and $93$ K for $\delta=0.95$ of Y123. }
\label{tab:Hall}
\begin{ruledtabular}
\begin{tabular}{lllllll}
 & Sample & $T_{c}$ (K) & Doping ($p$) & $A_H$ & $B_H$ &
$C_H$  \\
\hline
La214 & 0.21 & 30.99 & 0.21  & 0.10 & 0.52 & 0.34\\
 & 0.18 & 37.14 & 0.18  & 0.25 & 1.08 & 0.23\\
 & 0.15 & 37.93 & 0.15  & 0.54 & 1.40 & 0.035\\
 & 0.12 & 31.88 & 0.12  & 0.96 & 1.62 & -0.42\\
 & 0.08 & 22.3 & 0.08  & 2.23 & 2.15 & -2.04\\
 & 0.05 & 2.66 & 0.05  & 4.34 & 2.68 & -5.62\\
% & Hwang & &  & 0.23 & 0.0022 & -0.065\\
La-Bi2201 & 0.24 & 24.06 & 0.218 & 0.14 & 0.91 & 0.91 \\
 & 0.3 & 30.0 & 0.194  & 0.22 & 0.87 &  1.11\\
 & 0.44 & 33.03 & 0.16  & 0.35 & 1.12 & 1.00\\
 & 0.57 & 28.44 & 0.118  & 0.49 & 1.23 & 1.08\\
 & 0.66 & 19.95 & 0.09  & 0.61 & 1.27 & 1.70 \\
 Y123 & 0.55 & 55.0 & 0.088  &1.76 & 1.27 & 0.56 \\
 & 0.6 & 57.0 & 0.091  & 1.52 & 1.15 &  0.20\\
 & 0.65 & 58 & 0.092  & 1.46 & 1.09 & -0.035\\
 & {\bf 0.7} & {\bf 60.0} & {\bf 0.094} & {\bf 1.0} & {\bf 1.0} & {\bf 0.0} \\
 & 0.75 & 62.0 & 0.096  & 0.86 & 1.05 &  -0.20\\
 & 0.8 & 69.0 & 0.103  & 0.67 & 1.08 & -0.11\\
 & 0.85 & 83.0 & 0.123  & 0.70 & 0.96 & -0.17\\
 & 0.95 & 93.0 & 0.16  & 0.35 & 0.78 & 0.073\\
 & 7.0 & 91.0 & 0.176  & 0.28 & 0.67 & 0.031
\end{tabular}
\end{ruledtabular}
\end{table}

To reveal the physics behind the anomalous temperature dependence
of the Hall effect without invoking a specific model, we have
analyzed the scaling behavior of $R_H$. We assume the scaling
function of $R_H$ to have the form defined as in Eq.
(\ref{sec2:eq1b}):
\begin{equation}
R_{H}(T) = A_H f_H \left( \frac{T}{B_H} \right) + C_H.
\label{eq:Hall2}
\end{equation}
By fitting the experimental data with this formula using the method
introduced in Sec.~\ref{sec:2}, the scaling parameters $(A_H, B_H,
C_H)$ and the scaling function $f_H$ can then be determined.

Figure~\ref{fig:Hall1} shows the scaling function $f_H$ for Y123,
La214, and La-Bi2201. The corresponding scaling parameters are
given in Table \ref{tab:Hall}. The experimental data were
extracted from Ref.~\onlinecite{segawa:104521} for Y123, from
Ref.~\onlinecite{ono:024515} for La214, and from
Ref.~\onlinecite{ando:R6991} for La-Bi2201. The experimental data
of Y123 with $\delta < 0.55$ are not included since the
temperature range measured is too narrow to allow a reliable
scaling analysis to be done.

For La214, we find that $R_H$ exhibits a good scaling behavior in
nearly the whole temperature range, as shown in
Fig.~\ref{fig:Hall1}(a). This is also true for La-Bi2201
[Fig.~\ref{fig:Hall1}(b)]. It suggests that the dynamical behavior
of $R_H$ is still governed by a single energy scale. However, in
contrast to the in-plane resistivity, the scaling curve of La214
cannot be perfectly scaled onto the scaling curve of La-Bi2201,
except in an intermediate temperature regime. For Y123, the
experimental data can also be scaled onto a single curve in the
high-temperature regime, above the peak temperature of $R_H$.
However, at low temperatures, $f_H$ shows very different
temperature dependence for different dopings. This difference
might be caused by the contribution of CuO chains in Y123.
Electrons in CuO chains are more disordered in the underdoped
samples than in the optimally doped one.

A similar scaling analysis has been done by Hwang \textit{et
al}.\cite{PhysRevLett.72.2636} and by Chen \textit{et
al}.\cite{PhysRevB.50.16125} The scaling equation they used is
essentially the same as Eq. (\ref{eq:Hall2}), but with different
notations. In their notation, the scaling equation is given by
\begin{equation}
R_{H}(T) = R_{H}^\infty +R_{H}^* f\left(\frac{T}{T^*} \right).
\label{eq:Hall1}
\end{equation}
$R_H^\infty$ is the high-temperature value of $R_H$, which is
approximately temperature independent at high temperatures. $T^*$
is a characteristic temperature to be determined. Empirically,
they assumed $T^*$ to be the crossover temperature from a
temperature dependent to a temperature independent $R_H$ at high
temperature. However, the crossover temperature (if exists) is
very high, well above the temperature range they measured, in the
underdoped samples. They cannot determine reliably the crossover
temperature, even by extrapolation. Thus, their scaling analysis
cannot be applied to the underdoped samples. This is not a problem
in our approach. In Fig.~\ref{fig:Hall1}(a), we compared the
scaling curve obtained by Hwang \textit{et
al}.\cite{PhysRevLett.72.2636} with ours. We find that these two
scaling curves agree well with each other for La214 above the peak
temperature of $R_H$. This suggests that the crossover temperature
they determined is proportional to the parameter $B_H$ as we
determined here.

Recently, Gor'kov and Teitel'baum\cite{gor'kov:247003} and Ono
\textit{et al}.\cite{ono:024515} analyzed the high-temperature
behavior of $R_H$ using a two-band model in La214. They assumed
the high-temperature data of $R_H$ to be thermally activated,
resulting from strong charge fluctuations between the effective
lower and upper Hubbard bands. They found that the
high-temperature data of $R_H$ can indeed by explained by this
simple picture. Their results suggest that the thermal excitation
gap between the lower and upper Hubbard bands is significantly
smaller than the (direct) optical charge transfer gap.

\begin{widetext}
\begin{center}
\begin{figure}[tbp]
\includegraphics[width=16cm]{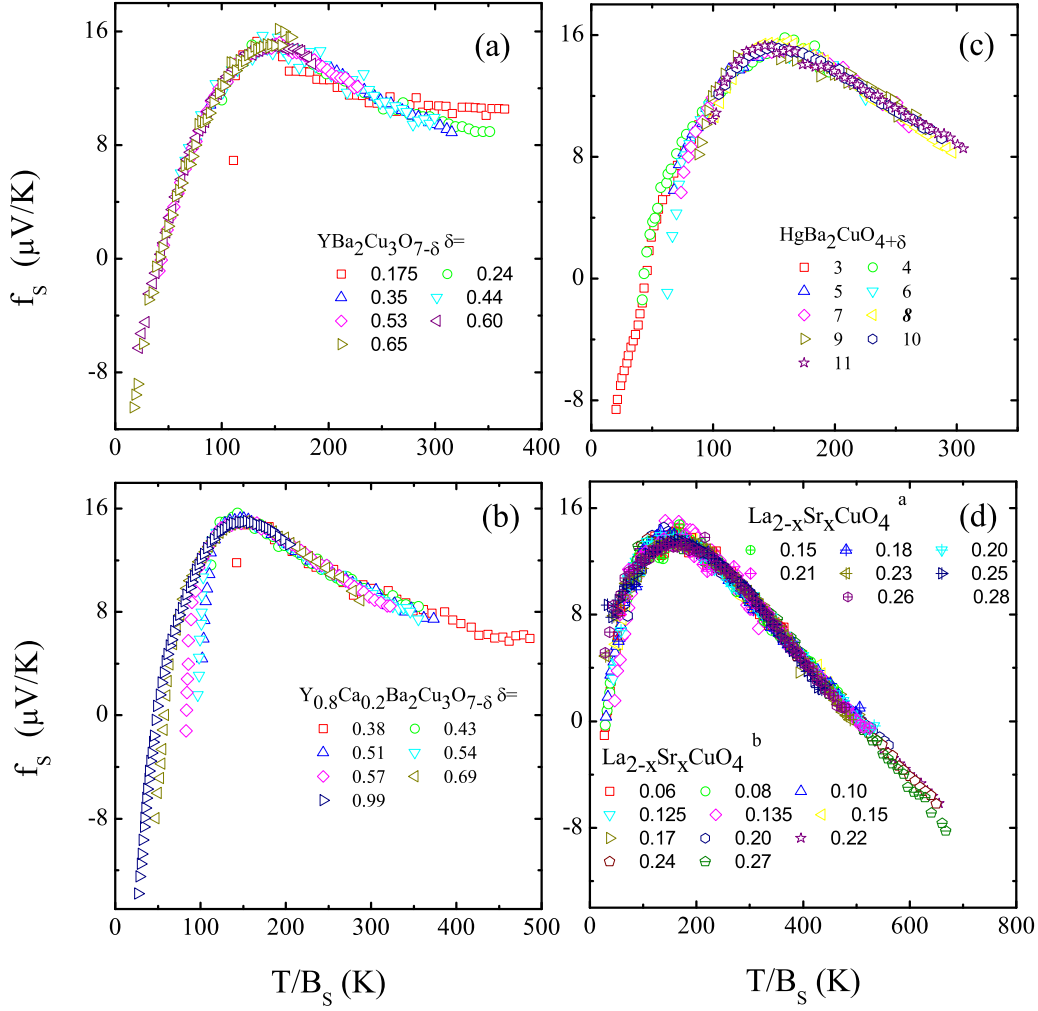}
\caption{The scaling function of the thermoelectric power $S(T)$
for Y123 (Ref.~\onlinecite{JPhys6.2237}), Ca-Y123
(Ref.~\onlinecite{PhysRevB.54.10201}), Hg1201
(Ref.~\onlinecite{PhysRevB.63.024504}), La214$^{a}$
(Ref.~\onlinecite{PhysRevB.51.3104}), and La214$^{b}$
(Ref.~\onlinecite{JPhys6.2237}). The notations defined in
Ref.~\onlinecite{PhysRevB.63.024504} are used to label the Hg1201
samples. Sample $8$ of Hg1201 is the reference.} \label{fig:tep1}
\end{figure}
\end{center}
\end{widetext}

\subsection{Thermoelectric power $S(T)$} \label{TEP}

The thermoelectric power or the Seebeck coefficient $S(T)$ is one
of the transport quantities complementary to the resistivity and
Hall effect. It reveals the properties of quasiparticle
excitations both near and away from the Fermi level. It can be
used to judge whether the charge carriers are electrons or holes
from the sign of $S(T)$. It can also be used to quantify the
charge carrier concentration. In HTSC, empirically, the value of
$S(T)$ at $T=290$ K was found to be a good measure of the hole
concentration.\cite{PhysRevB.46.14928}

In high-$T_c$ oxides, the thermopower $S(T)$ is small in the
superconducting state due to the suppression of the pairing gap to
the quasiparticle excitations.\cite{PhysRevB.46.14928} In the
normal state, $S(T)$ increases with temperature. It exhibits a
maximum and then drops monotonically at high temperatures. At a
given temperature, $S(T)$ decreases with increasing doping. For
most of the high-$T_c$ compounds, including
Bi2212,\cite{PhysRevB.46.14928,JPC8.3047,PhysRevB.57.7491,JCP12.6199}
La-Bi2201,\cite{PhysRevB.62.622, PhysRevB.66.020503}
Tl1212,\cite{PhysRevB.46.14928,JPC8.3047} and
Hg1201,\cite{PhysRevB.63.024504, PhysRevB.65.104505, honma:214517}
$S(T)$ is positive in the underdoped regime, but becomes negative
in the overdoped regime. At optimal doping, $S(T)$ becomes
negative at high temperature. However, in La214, $S(T)$ was found
to be positive in the whole doping range.\cite{PhysRevB.51.3104,
JPhys6.2237, PhysRevB.59.1491} The thermopower of overdoped
Y-based cuprates also behaves differently from other compounds. It
shows a positive slope, in contrast to the negative slope in other
compounds, at high temperature. This is probably due to the
contribution of CuO chains.

\begin{table}[ht]
\caption{The scaling parameters for the thermoelectric power
$S(T)$ of Y123, Ca-Y123, and Hg1201 in Fig. \ref{fig:tep1}. The
unit of $C_S$ is $\mu$V/K. Sample $8$ of Hg1201 is taken as a
reference.} \label{tab:tep1}
\begin{ruledtabular}
\begin{tabular}{lllllll}
 & Sample & $T_{c}$ (K) & Doping (p)  & $A_S$ & $B_S$ & $C_S$ \\
 \hline
Y123 & 0.65 & &   & 75.53 & 1.80 & 4.17\\
 & 0.60 & 13.31 & 0.057  & 60.37 & 1.60 & 3.22\\
 & 0.53 & 44.66 & 0.079  & 42.42 & 1.27 & 2.46\\
 & 0.44 & 55.51 & 0.089  & 15.67 & 0.99 & 2.36\\
 & 0.35 & 59.8 & 0.093  & 10.36 & 0.94 & 2.03\\
 & 0.24 & 74.4 & 0.109  & 1.18 & 0.83 & 1.46\\
 & 0.175 & 90.7 & 0.14  & -1.46 & 0.79 & 0.48\\
Ca-Y123 & 0.38 & 85.50 & 0.16  & 0.31 & 0.61 & -1.17\\
 & 0.43 & 85.00 & 0.151  & 0.41 & 0.80 & -1.07\\
 & 0.51 & 81.58 & 0.136  & 0.99 & 0.81 & -1.57\\
 & 0.54 & 78.55 & 0.128  & 1.10 & 0.84 & 1.13\\
 & 0.57 & 77.19 & 0.125  & 1.15 & 0.85 & 4.21\\
 & 0.69 & 47.04 & 0.085  & 1.46 & 1.05 & 12.86\\
 & 0.99 & 37.69 & 0.077  & 2.01 & 1.40 & 30.07 \\
 Hg1201: &3 &  26.0 & 0.05 &3.65 & 1.58 &33.56\\
&4 &  46.0 & 0.057  & 3.81 & 1.43 & 11.79\\
&5 &  62.0 & 0.069  & 2.50 & 1.33 & 15.24\\
&6 &  72.0 & 0.09 & 1.98 & 1.26 & 4.88 \\
&7 &  77.0 & 0.103  & 1.90 & 1.12 & -3.27\\
&{\bf 8} &  {\bf 83.0} & {\bf 0.11} & {\bf 1.0} & {\bf 1.0} & {\bf 0.0} \\
&9 &  91.0 & 0.119  & 1.18 & 1.07 & -4.67\\
&10 &  95.0 & 0.127  & 0.96 & 1.01 & -2.80\\
&11 &  98.0 & 0.157  & 0.68 & 0.98 & -3.01\\
\end{tabular}
\end{ruledtabular}
\end{table}

\begin{table}[ht]
\caption{The scaling parameters for the thermoelectric power
$S(T)$ of La214 in Fig. \ref{fig:tep1}. The data are referenced to
sample $8$ of Hg1201 shown in Table \ref{tab:tep1}. The unit of
$C_S$ is $\mu$V/K. T$_{c}$ of La214$^{a}$ is obtained from the
data published in Ref.~\onlinecite{loram:1999} by interpolation.
$T_c$ of La214$^{b}$ is obtained from Fig. 1 of
Ref.~\onlinecite{JPhys6.2237}. } \label{tab:tep2}
\begin{ruledtabular}
\begin{tabular}{lllllll}
 & Sample & $T_{c}$ (K) & Doping (p) & $A_S$ & $B_S$ & $C_S$ \\
 \hline La214$^a$ & 0.15 & 37.93 & 0.15  & 1.23 & 0.87 & 16.44\\
 & 0.18 & 37.14 & 0.18  & 0.90 & 0.69 & 8.07\\
 & 0.2 & 34.62 & 0.2  & 0.74 & 0.63 & 5.70\\
 & 0.21 & 30.99 & 0.21  & 0.64 & 0.60 & 4.46\\
 & 0.23 & 23.77 & 0.23  & 0.52 & 0.61 & 1.95\\
 & 0.25 & 17.29 & 0.25  & 0.45 & 0.65 & 0.15\\
 & 0.26 & 14.4 & 0.26  & 0.42 & 0.72 & -1.48\\
 & 0.28 & 8.08 & 0.28  & 0.44 & 0.67 & -0.85\\
La214$^b$ & 0.06 & 5.0 & 0.06  & 53.51 & 2.01 & 4.18\\
 & 0.08 & 22.3 & 0.08  & 51.78 & 1.74 & 3.22\\
 & 0.1 & 30.75 & 0.1  & 40.03 & 1.54 & 2.53\\
 & 0.125 & 32.0 & 0.125  & 31.55 & 1.17 & 1.66\\
 & 0.135 & 35.9 & 0.135  & 30.04 & 1.01 & 0.88\\
 & 0.15 & 37.93 & 0.15  & 18.66 & 0.78 & 0.79\\
 & 0.17 & 38.4 & 0.17  & 17.73 & 0.69 & 0.87\\
 & 0.2 & 34.62 & 0.2  & 10.59 & 0.62 & 0.83\\
 & 0.22 & 27.81 & 0.22  & 6.34 & 0.54 & 0.61\\
 & 0.24 & 19.96 & 0.24  & 3.41 & 0.54 & 0.53\\
 & 0.27 & 12.0 & 0.27  & 2.69 & 0.53 & 0.43
\end{tabular}
\end{ruledtabular}
\end{table}

We have applied our scaling approach to the thermoelectric power.
The scaling formula is given by
\begin{equation}
S(T) =  A_S f_S \left( \frac{T}{B_S} \right) + C_S .
\label{eq:TEP7}
\end{equation}
Figure.~\ref{fig:tep1} shows the scaling curves for Y123, Ca-Y123,
Hg1201, and La214. The scaling parameters are listed in Tables
\ref {tab:tep1} and \ref{tab:tep2}. The experimental data were
extracted from Ref. \onlinecite{JPhys6.2237} for Y123, from Ref.
\onlinecite{PhysRevB.54.10201} for Ca-Y123, from Ref.
\onlinecite{PhysRevB.63.024504} for Hg1201, and from
Refs.~\onlinecite{PhysRevB.51.3104} and \onlinecite{JPhys6.2237}
for La214. The scaling curves for two heavily underdoped samples
with $T_{c}<2$ K and three heavily overdoped samples of Hg1201
(Ref.~\onlinecite{PhysRevB.63.024504}) deviate significantly from
the universal scaling curve and are not included in the figure.
For Y123 and Ca-Y123, we only analyze the underdoped samples since
in the overdoped regime the chain contribution becomes important,
which breaks the scaling law. The chain contribution can, in fact,
be seen already in the slightly underdoped sample of Y123 ($\delta
= 0.175$), whose high-temperature data of $S(T)$ already begin to
deviate away from the scaling curve at high temperatures. For some
samples of Hg1201 and Ca-Y123, the measurement data fall faster
than the universal scaling curves at low temperatures. This can be
attributed to the superconducting fluctuations.

The scaling behavior of $S(T)$ in HTSC has been extensively
studied by a number of groups using the scaling formula like that
defined in Eq.~(\ref{sec2:sca3}).\cite{JPC8.3047, JPhys6.2237,
JCP12.6199, PhysRevB.63.024504, honma:214517} Our scaling curves
are consistent with their results. However, our data are much less
scattered than theirs.

\subsection{Uniform magnetic susceptibility $\chi(T)$}
\label{Chi}

The uniform magnetic susceptibility $\chi(T)$ measures basically
the density of states at the Fermi level in conventional Landau
Fermi liquid. However, in high-$T_c$ copper oxides, the
susceptibility is strongly affected by antiferromgantic spin
fluctuations. The parent compounds of HTSC are half-filled
antiferromagnetic Mott insulators with long-range N\'eel
order.\cite{PhysRevLett.58.2802} Upon doping, the N\'{e}el order
is rapidly suppressed, but antiferromagnetic fluctuations persist
up to slightly overdoping.\cite{wakimoto:247003,wakimoto:217004}

At half-filling, $\chi(T)$ shows a sharp peak around the N\'eel
temperature.\cite{PhysRevB.38.905, PhysRevB.38.6636} This peak
shifts down to lower temperature with doping and disappears
completely when the superconductivity emerges. In the normal
state, the magnetic susceptibility $\chi(T)$ first increases with
increasing temperature, develops a broad peak, and then drops down
at high temperature.\cite{PhysRevB.40.8872, PhysRevB.40.2254,
PhysRevLett.62.957, PhysRevB.41.2605, PhysRevB.48.9747,
PhysRevB.49.16000, PhysRevLett.84.5848}

In real materials, the magnetic susceptibility is strongly
affected by magnetic impurities. These impurities contribute a
Curie term to $\chi$, which diverges as $1/T$ at low temperatures.
The impurity contribution is strongly sample dependent. In order
to analyze the intrinsic behavior of the magnetic susceptibility,
this Curie term of impurities should be subtracted from the raw
data first.

\begin{figure}[ht]
\includegraphics[width=9cm]{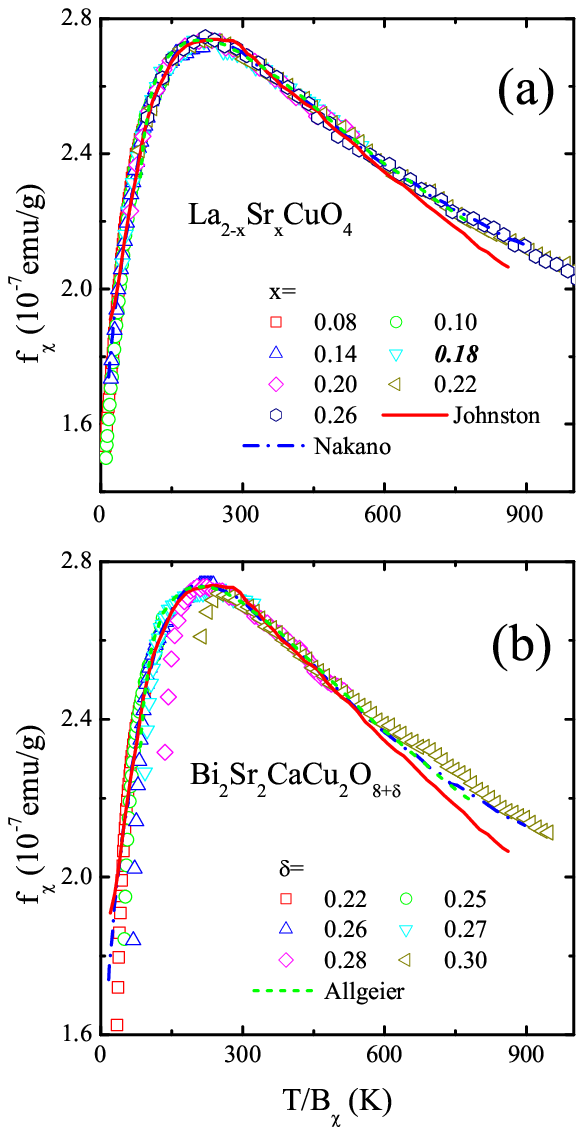}
\caption{The scaling function of the uniform magnetic
susceptibility $\chi(T)$ for La214
(Ref.~\onlinecite{PhysRevB.49.16000}) and Bi2212
(Ref.~\onlinecite{PhysRevLett.84.5848}). The solid, dash-dot and
dash lines represent the rescaled scaling curves obtained by
Johnston (Ref.~\onlinecite{PhysRevLett.62.957}), Nakano \textit{et
al} (Ref.~\onlinecite{PhysRevB.49.16000}) and Allgeier and
Schilling (Ref.~\onlinecite{PhysRevB.48.9747}), respectively.}
\label{fig:chi1}
\end{figure}

\begin{figure}[ht]
\includegraphics[width=9cm]{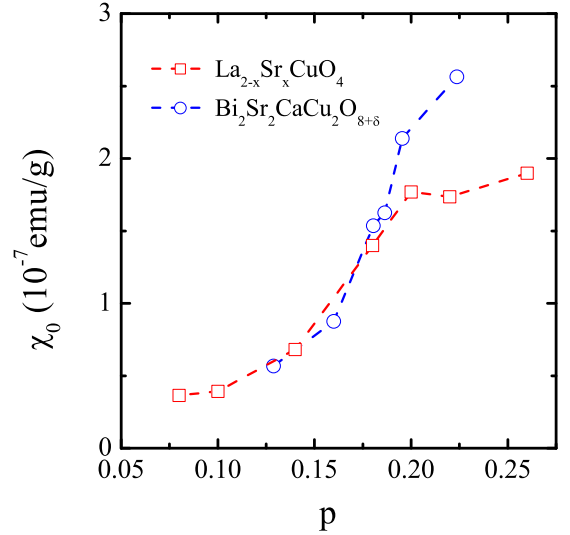}
\caption{The temperature independent magnetic susceptibility
$\chi_{0}=C_{\chi}+A_{\chi} \chi_{0,s}$ where
$\chi_{0,s}=1.4\times 10^{-7}$emu/g for La214 with $x=0.18$ as
obtained from Nakano \textit{et al}.
(Ref.~\onlinecite{PhysRevB.49.16000}). } \label{fig:chi3}
\end{figure}

\begin{table}[h]
\caption{The scaling parameters for the magnetic susceptibility
$\chi(T)$ shown in Fig.~\ref{fig:chi1}. The unit of $C_{\chi}$ is
$10^{-7}$ emu/g. $T_{c}$ of La214 is obtained from
Ref.~\onlinecite{loram:1999} by interpolation. In obtaining the
scaling curves, a Curie term $C/T$ is subtracted from the
measurement data for the La214 samples with $x=0.20, 0.22$, and
$0.26$. The corresponding values of $C$, obtained from Fig. 8 of
Ref.~\onlinecite{PhysRevB.49.16000}, are $12.41, 37.23$, and
$60.23$ $10^{-7}$ emu/g, respectively. } \label{tab:chi}
\begin{ruledtabular}
\begin{tabular}{lllllll}
 & Sample & $T_{c}$ (K) & Doping (p)  & $A_{\chi}$ &$B_{\chi}$ &$C_{\chi}$\\
\hline
 La214 & 0.08 & 22.3 & 0.08  & 1.40 & 5.97 & -1.60\\
 &0.10 & 30.75 & 0.10  & 1.41 & 4.18 & -1.57\\
 & 0.14& 36.62 & 0.14  & 1.28 & 2.31 & -1.11\\
 & {\bf 0.18} & {\bf 37.45} & {\bf 0.18} & {\bf 1.0} & {\bf 1.0} & {\bf 0.0} \\
 & 0.20 & 34.62 & 0.20  & 1.04 & 0.68 & 0.31\\
 & 0.22  & 27.81 & 0.22  & 1.24 & 0.53 & -0.001\\
 & 0.26 & 14.57 & 0.26  & 1.38 & 0.46 & -0.027\\
% & Johnston & &   & 0.64 & 0.0043 & -0.76\\
% & Allgeier & &  & 0.056 & 0.0075 & -0.11\\
% & Nakano &  &   & 0.71 & 0.0044 & -0.96\\
 Bi2212 & 0.22 & 82.0 & 0.13  & 1.33 & 3.04 & -1.29\\
 & 0.25 & 89.57 & 0.16  & 1.10 & 2.22 & -0.66\\
 & 0.26 & 87.19 & 0.18  & 0.66 & 1.60 & 0.61\\
 & 0.27 & 83.45 & 0.186  & 0.72 & 1.18 & 0.62\\
 & 0.28 & 78.7 & 0.195  & 0.51 & 0.74 & 1.42\\
 & 0.30 & 68.51 & 0.22  & 0.66 & 0.40 & 1.84
\end{tabular}
\end{ruledtabular}
\end{table}

To elucidate the intrinsic property of the magnetic
susceptibility, we have analyzed the scaling behavior of $\chi(T)$
with the following single-parameter scaling equation:
\begin{equation}
\chi(T) = A_{\chi} f_{\chi} \left( \frac{T}{B_{\chi}} \right) +
C_{_\chi}. \label{eq:chi2}
\end{equation}
Figure \ref{fig:chi1} shows the scaling curves of $\chi(T)$ for
La214 and Bi2212. The experimental data were extracted from those
published by Nakano \textit{et al.}\cite{PhysRevB.49.16000} for
La214 and by Watanabe \textit{et al.}\cite{PhysRevLett.84.5848}
for Bi2212. The corresponding scaling parameters are shown in
Table \ref{tab:chi}. In the scaling analysis for the La214 samples
with $x=0.20, 0.22$, and $0.26$, a Curie term $C/T$ is subtracted
from the experimental data; the corresponding values of $C$, i.e.,
$C = 12.41, 37.23$, and $60.23$ (in units of $10^{-7}$ emu/g),
were obtained by Nakano \textit{et al.}\cite{PhysRevB.49.16000}

We find that the susceptibility for both La214 and Bi2212 exhibits
a good scaling behavior. For Bi2212, $\chi(T)$ in heavily
overdoped samples begins to deviate from the scaling curve near
$T_c$. This is likely to be due to strong superconducting
fluctuations.

Our universal scaling curve is consistent with the scaling
analysis given by Johnston\cite{PhysRevLett.62.957} and Nakano
\textit{et al.}\cite{PhysRevB.49.16000} for La214, and by Allgeier
and Schilling\cite{PhysRevB.48.9747} for Bi2212. The scaling
analysis of Nakano \textit{et al.} was made based on the scaling
formula defined by Eq.~(\ref{sec2:sca3}).\cite{PhysRevB.49.16000}
The problem with that kind of analysis is that the characteristic
temperature $T^*$ defined in Eq.~(\ref{sec2:sca3}) has to be
determined empirically prior to the scaling analysis. The
characteristic temperature $T^*$ for the susceptibility was
generally determined from the peak temperature of $\chi$. However,
in heavily overdoped samples, no peak structure has been observed
within the whole temperature measured. This has limited the
application of that kind of scaling analysis. In addition, to
fully satisfy Eq.~(\ref{sec2:sca3}), a constant term also needs to
be subtracted for each set of data. This is also difficult if the
measured temperature range is not broad enough. Nevertheless, we
find that their scaling curves, as shown in Fig.~\ref{fig:chi1},
agree well with ours.

The scaling analyses given by Johnston\cite{PhysRevLett.62.957}
and by Allgeier and Schilling\cite{PhysRevB.48.9747} are based on
the high-temperature series expansion for a two-dimensional
antiferromagnetic Heisenberg model. The scaling function obtained
by Johnston deviates slightly from the universal scaling curve
obtained at high temperature. The high-temperature scaling curve,
as shown by Johnston\cite{PhysRevLett.62.957} and by Allgeier and
Schilling,\cite{PhysRevB.48.9747} agrees with the temperature
dependence of the susceptibility of the two-dimensional
antiferromagnetic Heisenberg model without doping.

In obtaining the scaling function, a temperature independent term
is subtracted from $\chi(T)$. This term can be expressed as
\begin{equation}
\chi_0 = C_{\chi} + A_{\chi} \chi_{0,s},
\end{equation}
where $\chi_{0,s}$ is the value of $\chi_0$ for the reference
sample. The variation of $\chi_0$ with doping may reflect the
change of the density of states at the Fermi level. Thus, it is
interesting to analyze the doping dependence of this term.

Figure~\ref{fig:chi3} shows the doping dependence of $\chi_0$.
$\chi_0$ increases monotonically with doping. This temperature
independent term might be the contribution of the core
diamagnetism, the Van Vleck paramagnetism, the Landau
diamagnetism, and the Pauli paramagnetism of the band electrons if
the effect of antiferromagnetic correlations is ignored. The core
diamagnetism and the Van Vleck paramagnetism are doping
independent. The Landau diamagnetism is generally small. Thus, the
doping dependence of $\chi_0$ is mainly affected by the Pauli
susceptibility, which is proportional to the density of states at
the Fermi level. Hence, the change of $\chi_0$ with doping will
correspond to the change of the density of states at the Fermi
level. This simple observation is consistent with the result of
Allgeier and Schilling\cite{PhysRevB.48.9747} as well as the
measurement of the specific heat.\cite{loram:1999}

\subsection{Spin-lattice relaxation rate $1/T_1$} \label{NMR}

The nuclear magnetic resonance probes the local spin dynamics via
the measurement of the Knight shift, the spin-lattice relaxation
rate $1/T_{1}$, and other spin response
functions.\cite{pennigton:1990} The Knight shift measures the
shift of the resonance frequency induced by the conduction
electrons. It is proportional to the uniform magnetic
susceptibility $\chi(T)$. The temperature dependence of the Knight
shift should follow the scaling law of the uniform susceptibility
as discussed in the previous section. This is, indeed, supported
by the experimental measurement (see, for example, Ref.
\onlinecite{barzykin:247002}). Below we will discuss the scaling
behavior of the spin-lattice relaxation rate $1/T_1$.

The high-temperature dependence of $1/T_{1}T$ of $^{63}$Cu shows a
Curie-Weiss-like behavior.\cite{PhysRevB.54.545,PhysRevB.54.10131}
This can be attributed to the contribution of antiferromagnetic
fluctuations since $1/T_{1}T$ of $^{63}$Cu is dominated by spin
fluctuations near $Q = (\pi, \pi)$. It exhibits a broad maximum in
an intermediate temperature regime and then drops at low
temperature.

\begin{figure}[ht]
\includegraphics[width=9cm]{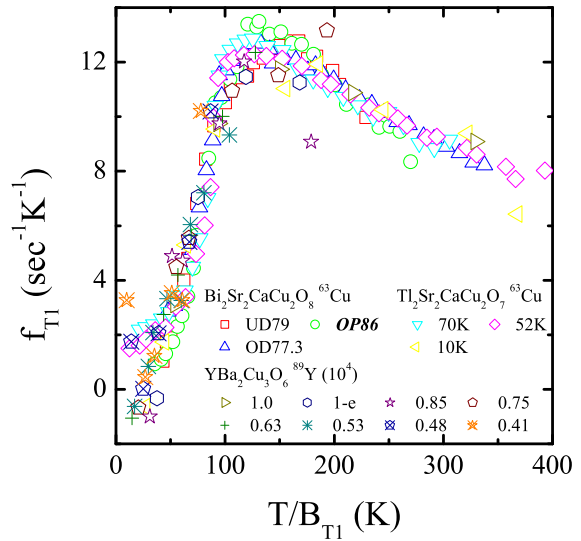}
\caption{The scaling function for $1/T_{1}T$ of $^{63}$Cu in
Bi2212 and Tl1212, and that of $^{89}$Y in Y123. The experimental
data are obtained from Ref.~\onlinecite{PhysRevB.58.R5960} for
Bi2212, from Ref.~\onlinecite{PhysRevB.54.10131} for Tl1212, and
from Ref.~\onlinecite{PhysRevLett.63.1700} for Y123. The notations
of the samples are the same as those in the corresponding
references. The Bi2212 $OP86$ is taken as the reference sample.}
\label{fig:NMR1}
\end{figure}

\begin{table}[h]
\caption{The fitting parameters, $A_{T1}$, $B_{T1}$, and $C_{T1}$
for the scaling analysis of $1/T_{1}T$ shown in
Fig.~\ref{fig:NMR1}. The unit of $C_{T1}$ for $^{63}$Cu of Bi2212
and Tl1212 is s$^{-1}$K$^{-1}$, and for $^{89}$Y of Y123 is
$10^{-4}$ s$^{-1}$K$^{-1}$. } \label{tab:NMR}
\begin{ruledtabular}
\begin{tabular}{lllllll}
 & Sample & $T_c$ (K) & Doping (p)  & $A_{T1}$ & $B_{T1}$ & $C_{T1}$\\
\hline Bi2212 $^{63}$Cu & 0.125 & 79.0 & 0.13  & 0.86 & 1.31 & 0.72\\
& {\bf 0.20} & {\bf 86.0} & {\bf 0.16} & {\bf 1.0} & {\bf 1.0} &
{\bf 0.0} \\
& 0.225 & 77.3 & 0.20 & 1.17 & 0.85 & -1.67\\
Tl1212 $^{63}$Cu & & 70.0 &  & 2.44 & 0.72 & -4.72\\
& & 54.0 &  & 2.29 & 0.56 & -3.05\\
& & 10.0 &  & 0.48 & 0.16 & 16.18\\
Y123 $^{89}$Y & 1.0 & 88.8 & 0.19  & 0.075 & 0.88 & 1.51\\
& 1-$\varepsilon$ & &  & 0.069 & 1.35 & 1.53\\
& 0.85 & 90.3 & 0.14  & 0.085 & 1.63 & 1.15\\
& 0.75 & 67.9& 0.10  & 0.087 & 1.50 & 0.83\\
& 0.63 & 57.3 & 0.09  & 0.088 & 2.30 & 0.90\\
& 0.53 & 52.5 & 0.086  & 0.097 & 2.80 & 0.80\\
& 0.48 & 38.0 & 0.07  & 0.070 & 3.35 & 0.77\\
& 0.41 & 15.0 & 0.06  & 0.028 & 3.74 & 0.68
\end{tabular}
\end{ruledtabular}
\end{table}

We have analyzed the scaling behavior of the experimental data of
$1/T_{1}T$ published by Ishida \textit{et
al}.\cite{PhysRevB.58.R5960} for $^{63}$Cu in Bi2212, by Magishi
\textit{et al}.\cite{PhysRevB.54.10131} for $^{63}$Cu in Tl1212,
and by Alloul \textit{et al}.\cite{PhysRevLett.63.1700} for
$^{89}$Y in Y123. The scaling equation is assumed to be
\begin{equation}
\frac{1}{T_{1}T} = A_{T1} f_{T1}\left(\frac{T}{B_{T1}}\right) +
C_{T1}. \label{eq:NMR1}
\end{equation}
Figure~\ref{fig:NMR1} shows the temperature dependence of the
scaling function $f_{T1}$. The corresponding scaling parameters
are given in Table.~\ref{tab:NMR}.

Within experimental errors, we find that $1/T_{1}T$ shows a good
scaling behavior. For all the materials shown in
Fig.~\ref{fig:NMR1}, the data of $1/T_{1}T$ can be scaled on a
common curve in a relatively wide range of temperature. This,
again, suggests that the normal state dynamics is controlled by a
single energy scale. This single-parameter scaling behavior is
consistent with the existing theory of spin fluctuations, such as
the antiferromagnetic Fermi liquid theory proposed by Millis
\textit{et al}. \cite{PhysRevB.42.167}

\section{Analysis of the characteristic energy scale} \label{sec:4}

In the preceding section, we have analyzed the scaling behaviors
of the transport coefficients, including the resistivity, the Hall
effect, and the thermoelectric power, and the magnetic response
functions, including the spin susceptibility and the spin-lattice
relaxation rate, in the normal state of HTSC. These coefficients
probe different aspects of low-energy excitations and are
physically distinct. However, we find that they all show good
scaling behaviors. For most of the measurement quantities, the
temperature dependence of the corresponding scaling functions are
universal, depending neither on the doping concentration nor on
the chemical structure of the materials measured.

Among the three scaling parameters, the energy scale $\Delta$ or
the relative energy scale $B$ defined in Eq.~(\ref{sec2:eq5}) is
the most important one. It characterizes the basic energy scale
governing the temperature dependence of a response function. For
the c-axis resistivity, we have determined the absolute values of
$\Delta$ using the approximate scaling function of $\rho_c$
derived in our previous work for multilayer
cuprates.\cite{su:134510} For other measurement quantities, as the
analytic formula of the scaling functions are unknown, only the
ratio of $\Delta$ with respect to a reference sample, $B=\Delta /
\Delta_s$, is determined. Nevertheless, we find that these
characteristic energy parameters determined from different
coefficients show a common trend with doping. As shown in
Tables~\ref{tab:rhoc2}-\ref{tab:tep2}, they all decrease with
increasing doping.

\begin{figure}[ht]
\includegraphics[width=9cm]{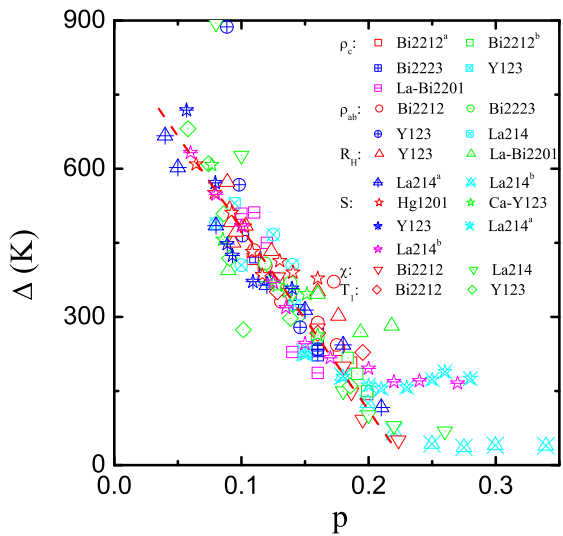}
\caption{The doping dependence of the energy scales obtained with
Eq. (\ref{rescale-delta}). The values of the scaling factor
$\eta_y$ are listed in Table \ref{tab-rescale}. The dashed line is
the linear fit to the scaled pseudogap energies obtained by
various experimental probes. Here, we take the pseudogap energy
obtained from the c-axis resistivity of Bi2212 as a reference, as
shown in Fig. \ref{fig:deltarc}. } \label{delta-p}
\end{figure}

\begin{table}[ht]
\caption{The values of the scaling factor $\eta_y$ for different
probes $y = \rho_c, \rho_{ab}, R_H, \chi, 1/T_1T$, and $S$ of the
cuprates. The superscripts $a$ and $b$ for $S$ and $R_{H}$ of
La214 refer to the data published in Ref.~\onlinecite{ono:024515}
and Ref.~\onlinecite{PhysRevLett.72.2636}, respectively. }
\label{tab-rescale}
\begin{ruledtabular}
\begin{tabular}{lcccccc}
  & $\rho_c$ & $\rho_{ab}$  & $R_H$ & $S$ & $\chi$ & $T_{1}T$ \\
 \hline
 Bi2212  & 1 & 455 & & & 125 &267 \\
 Bi2223  & 1.04 & 375 & & & & \\
 Y123  & 1.67 & 465  & 450 & 450 &  & 182 \\
 Ca-Y123 & &   & & 435 & &  \\
 La214  &  & 405 & 225$^a$ & 255$^a$  & 150 &\\
        & &    & 0.46$^b$ & 315$^b$  &   &    \\
 La-Bi2201& 450 &  & 310&  &  & \\
 Hg1201 & &  &  & 385 &  &
\end{tabular}
\end{ruledtabular}
\end{table}

The similar doping dependence of the characteristic energy
parameters suggests that these relative energy scales obtained by
different probes may have a common physical origin. This can be
examined by rescaling all the relative energy scales with respect
to the absolute energy scale of the pseudogap $\Delta$ obtained
from the scaling analysis of $\rho_c$. They should fall onto a
single curve if they are, indeed, the pseudogap energies. To do
this, let us introduce the following formula:
\begin{equation} \label{rescale-delta}
\Delta_y(p) = \eta_{y} B_{y}(p),
\end{equation}
where $\eta_y$ is a scaling factor and $\Delta_y$ is the
characteristic energy scale. The subscript $y$ represents the
measured physical quantity, i.e., $y= \rho_c,\, \rho_{ab}, \,R_H,
\, S,\, \chi, $ and $1/T_1T$. The scaling factors $\eta_{y}$ can
be determined by the least square fit using the approach
introduced in Sec.~\ref{sec:2}.

The fitting parameters of $\eta_y$ are given in
Table~\ref{tab-rescale}. By substituting them into
Eq.~(\ref{rescale-delta}), we can obtain the values of $\Delta_y$.
The result, as shown in Fig.~\ref{delta-p}, indicates that all the
energy scales determined from the scaling analysis given in
Sec.~\ref{sec:3} have the same doping dependence within
experimental errors, which result mainly from the uncertainty in
the determination of doping concentration. It suggests that all
the dynamic coefficients analyzed in Sec.~\ref{sec:3} are, indeed,
governed by the same energy scale. This is a remarkable result
since different coefficients probe different responses of charge
and/or spin degrees of freedom. For example, the c-axis
resistivity $\rho_c$ is susceptive to the charged excitations
around the antinodal points, while the in-plane resistivity
$\rho_{ab}$ is mainly affected by the scattering of charged
quasiparticles around the node points. The uniform magnetic
susceptibility probes the spin fluctuations around
$\mathbf{k}=(0,0)$, while the spin-lattice relaxation rate is
strongly dependent on the antiferromagnetic fluctuations around
$\mathbf{k}=(\pi,\pi)$.

In the underdoped regime, the characteristic energy $\Delta$ drops
almost linearly with doping. This is consistent with the doping
dependence of the pseudogap observed by
ARPES,\cite{Campuzano:1999} tunneling,\cite{Hufner:2007} and other
measurements.\cite{tacon:2006} Thus, the control energy scale in
this regime is, indeed, the pseudogap. If we extrapolate the
underdoped data of $\Delta$ to zero doping, we find that $\Delta$
is in order of $J$, and to higher doping, $\Delta$ vanishes
roughly at $p \sim 0.23$-$0.25$.

In the literature, two kinds of pseudogaps (or the onset
temperatures of pseudogap), which were often quoted as the
``large" and ``small" pseudogaps, were reported.\cite{timusk:1999}
The large pseudogap generally refers to the characteristic
temperature $T^*$ measured, for example, by the magnetic
susceptibility,\cite{PhysRevB.49.16000} the Hall
coefficient.\cite{PhysRevLett.72.2636}  The small one could be the
energy scale probed by other experimental techniques, such as the
transport properties measurements \cite{PhysicaC235.130} or the
leading-edge shift around the antinodal direction measured by the
ARPES.\cite{RevModPhys.75.473} Our scaling analysis indicates that
these two energy scales are, in fact, physically
indistinguishable.

In the overdoped regime, the universal scaling behavior generally
breaks down.  For overdoped La214, $\Delta_y$ determined from the
thermopower and the Hall coefficients separates into two branches
with different energies. $\Delta_y$ determined from the
thermopower is higher than that from the Hall effect. This may
imply the existence of two energy scales in the overdoped regime.
However, as the single-parameter scaling law still holds in this
regime for both the thermopower and the Hall coefficients, further
investigation on this issue is desired.

Recently, there is a surge of interest in the discussion of
two-gap energies, namely, the pseudogap in the normal state and
the relatively smaller superconducting gap in the superconducting
state. The existence of a distinct energy scale in the
superconducting state, whose doping dependence is different from
that of the pseudogap, was first reported in the penetration depth
measurements\cite{Panagopoulos:1998} and later in the Andreev
reflection measurements.\cite{Deustcher:1999}  This lower energy
scale characterizes the low-lying excitations around the gap nodes
and appears only in the superconducting state. Recently,
ARPES,\cite{tanaka:2006} Raman scattering,\cite{tacon:2006} and
inelastic neutron measurements\cite{Chang2007} have further
confirmed the existence of these two distinguished energy scales.
Furthermore, the ARPES has revealed that the two-gap structure is
intimately connected with the arc (or pocket) feature of the Fermi
surface of HTSC in the underdoped regime. It is believed that the
superconducting gap develops predominately on the Fermi arc below
$T_c$. This is consistent with the early ARPES measurement data.
\cite{PhysRevLett.83.840} However, there exists also other
experimental measurements, which suggest that there is only one
energy scale in the superconducting state and the superconducting
gap around the nodal points is nothing but an extension of the
pseudogap in the arc area in the superconducting
state.\cite{kanigel2007} This one energy scale scenario is
consistent with the picture of resonant valence bonds based on the
charge-spin separation\cite{PALee:2006} as well as that of
preformed pairs.\cite{Emery:1995}

In this work, we have shown that there is only one energy scale in
the normal state. However, as we have only analyzed the scaling
behavior of the experimental data in the normal state, we are unable
to address the issue of two energy gaps in the superconducting
state. It is of great interest to extend the single-parameter
scaling method introduced in Sec.~\ref{sec:2} to the superconducting
state at which two energy scales (or control parameters) may exist.
This would then allow us to judge whether there is only one or two
energy scales in the superconducting state from the
model-independent scaling analysis of various transport and
thermodynamic coefficients.

\section{Summary} \label{sec:5}

We have introduced a scaling method to study the scaling behavior
in the normal state of HTSC. We have analyzed the scaling behavior
of the c-axis resistivity, the in-plane resistivity, the Hall
coefficient, the thermoelectric power, the magnetic
susceptibility, and the nuclear magnetic resonance, and extracted
the corresponding energy scales. It is found that all these
quantities, no matter how different they are, exhibit universal
scaling behaviors, controlled by a single energy scale in the
normal state. Furthermore, we find that all these energy scales
obtained from different physical coefficients have the same doping
dependence as the pseudogap. It shows that the pseudogap is the
only characteristic energy governing the low-lying excitations in
the normal state of HTSC.

The scaling method we introduced in Sec.~\ref{sec:2} is model
independent. It provides a simple but powerful tool to analyze the
scaling behavior of experimental data. It can be applied not only to
the normal state of HTSC, but also to any other systems where the
single-parameter scaling hypothesis, i.e. Eq.~(\ref{sec2:eq1}) or
(\ref{sec2:eq1b}), is valid.

{\bf Acknowledgement} Support from the NSFC and the national
program for basic research of MOST of China is acknowledged.

%%%%%%%%%%%%%%%%%%%%%%%%%%%%%%%%%%%%%%%%%%%%%%%%%%%%%%%%%%%%%%%%%%%%%%%%%%%%%%%%%%%%%%%%%
%\bibliography{references}

\begin{thebibliography}{102}
\bibitem[{wat()}]{watanabe:note}
\bibinfo{note}{The experimental data of $\delta = 0.28$ in Bi2212 is from T.
  Watanabe.}
\end{thebibliography}
%%%%%%%%%%%%%%%%%%%%%%%%%%%%%%%%%%%%%%%%%%%%%%%%%%%%%%%%%%%%%%%%%%%%%%%%%%%%%%%%%%%%%%%%%

\end{document}